\documentclass[12pt]{article}

\usepackage{epsfig,amsmath}
\usepackage{amssymb}
\usepackage{cite}

\numberwithin{equation}{section}

\newcommand{\eg}{\emph{e.g.}\ }
\newcommand{\ie}{\emph{i.e.}\ }
\newcommand{\cfr}{\emph{cf.}\ }

\def\order#1{{\cal{O}}\left(#1\right)}

\def\as{\alpha_{{\textsc{s}}}}
\def\gae{{\gamma_{\textsc{e}}}}
\def\asb{{\bar \alpha}_{{\textsc{s}}}}
\newcommand{\mx}{\mathrm{max}}
\def\half{\mbox{\small $\frac{1}{2}$}}

\def\hc{{\cal H}_{\mathrm{C}}}
\def\hr{{\cal H}_{\mathrm{R}}}
\def\rc{{\cal R}_{\mathrm{C}}}
\def\rr{{\cal R}_{\mathrm{R}}}
\def\ec{{\cal E}_{\mathrm{C}}}

\def\vkti{\vec{k}_{ti}}
\def\vktj{\vec{k}_{tj}}
\def\vkt{\vec{k}_{t}}
\def\vpt{\vec{p}_{t}}

\def\vb{\vec{b}}

\def\kt{k_t}
\def\kti{k_{ti}}
\def\ktj{k_{tj}}

\def\vb{\vec{b}}

\def\Btilde{{\widetilde B}}
\def\Ltilde{{\widetilde L}}

\def\Covrln{{\overline C}}
\def\Govrln{{\overline G}}
\def\Lovrln{{\overline L}}
\def\Lovrlntilde{{\widetilde{\overline L}}}

\def\qbar{{\bar q}}
\def\Lbar{{\bar L}}
\def\LambdaBar{{\bar \Lambda}}
\def\Bbar{{\bar B}}
\def\Rpbar{{\bar R}'}
\def\Rppbar{{\bar R}''}
\def\Rpppbar{{\bar R}'''}
\newcommand{\nubar}{{\bar \nu}}
\newcommand{\bbar}{{\bar b}}
\newcommand{\zbar}{{\bar z}}

\newcommand{\bq}{\boldsymbol{q}}
\newcommand{\bP}{\boldsymbol{P}}

\newcommand{\bC}{\boldsymbol{C}}
\newcommand{\bg}{\boldsymbol{g}}
\newcommand{\bgamma}{\boldsymbol{\gamma}}

\def\al{\alpha}
\def\be{\beta}

\def\cM{{\cal{M}}}    
\def\cP{{\cal{P}}}    
\def\cI{{\cal{I}}}
\def\cJ{{\cal{J}}}
\newcommand{\cG}{\mathcal{G}}

\def\cR{{\cal{R}}}               

\def\half{\mbox{\small $\frac{1}{2}$}}

\def\MSbar{\overline{\mbox{\scriptsize MS}}}
\def\MSBar{\MSbar}

\def\eff{\mathrm{eff}}

\def\cf{C_F}
\def\CF{C_F}

\def\CA{C_A}
\def\nf{n_{\!f}}

\def\as{\alpha_{{\textsc{s}}}}
\def\gae{{\gamma_{\textsc{e}}}}
\def\asb{{\bar \alpha}_{{\textsc{s}}}}

\def\ee{e^+e^-}

\begin{document}

\begin{titlepage}
\begin{flushright}
{DESY--01--160 \\
  CERN--TH/2001--261\\
  LPTHE--01--43\\
  hep-ph/0110213 \\
  October 2001

}
\end{flushright}              
\vspace*{\fill}
\begin{center}
{\Large 
\textsf{\textbf{Resummation of the jet broadening in DIS}}\footnote{Research
    supported in part by the EU Fourth Framework 
    Programme `Training and Mobility of Researchers', Network `Quantum
    Chromodynamics and the Deep Structure of Elementary Particles',
    contract FMRX-CT98-0194 (DG 12-MIHT).}}
\end{center}
\par \vskip 5mm
\begin{center}
       
       {\large \textsf{ M.~Dasgupta}} \\ 
          DESY, Theory Group, Notkestrasse 85, Hamburg, Germany.
          \vspace{0.5cm}\\
       {\large \textsf{  G.~P.~Salam}} \\
  CERN, TH Division, 1211 Geneva 23, Switzerland.\\
  LPTHE, Universit\'es P. \& M. Curie (Paris VI) et Denis Diderot
  (Paris VII), Paris, France.
 
\end{center}
\par \vskip 2mm
\begin{center} {\large \textsf{\textbf{Abstract}}} \end{center}
\begin{quote}
  We calculate the leading and next-to-leading logarithmic resummed
  distribution for the jet broadening in deep inelastic scattering, as
  well as the power correction for both the distribution and mean
  value. A truncation of the answer at NLL accuracy, as is standard,
  leads to unphysical divergences. We discuss their origin and show
  how the problem can be resolved. We then
  examine DIS-specific procedures for matching to fixed-order
  calculations and compare our results to data. One of
  the tools developed for the comparison is an NLO parton
  distribution evolution code. When compared to PDF sets from MRST and
  CTEQ it reveals limited discrepancies in both.
\end{quote}
\vspace*{\fill}

\end{titlepage}
\setcounter{footnote}{0}

\section{Introduction}

In $\ee$ collisions there have been extensive studies of event shape
distributions involving comparisons to calculations which resum
logarithms at the edge of phase space
\cite{CTTW,CTW,JetRates,NewBroad,Cpar}. Much has been learnt from
these studies, for example precise determinations of $\as$ and strong
tests of recent novel approaches to hadronisation (see for example
\cite{eePower}), and even explicit measurements of the colour factors
of QCD \cite{eeColFacts}.

In the past few years the H1 and ZEUS experiments at HERA have
embarked on analogous studies of DIS current-hemisphere event shapes
\cite{H1OldData,H1NewData,H1ZeusConf}. Compared to most $\ee$ results,
a feature of the DIS measurements is that a wide range of $Q$ values
is probed by the same experiment. This is useful when one wishes to
isolate effects with a specific dependence on $Q$, such as the running
of $\as$ or hadronisation corrections. Additionally DIS event shapes
depend on radiation from the incoming proton, allowing one to study
non-trivial questions related to the process-independence of
hadronisation corrections and perhaps even issues such as intrinsic
transverse momentum.

With these motivations in mind we recently initiated a project to
resum a range of event shapes in DIS \cite{DIStauz,NonGlobal,DSDIS}. This
paper deals with the resummation in the $1+1$ jet limit, of the
distribution of the jet-broadening with respect to the photon axis,
$B_{zE}$, as measured in the current hemisphere of the Breit frame of
deep inelastic scattering and defined as
\begin{equation}
\label{eq:BzEDef}
B_{zE} = \frac{\sum_{i\in \hc} |\vec{p}_i \times \vec{n}|}{
  2\sum_{i\in \hc} |\vec{p}_i|}  \, .
\end{equation}
In the above equation $\vec{n}$ refers to the photon axis
(conventionally taken to be the $z$ axis) in the Breit frame and $\hc$
is the current hemisphere. The above definition is valid provided one
imposes a certain minimum energy cut-off $\ec >
\mathcal{E}_{\mathrm{lim}}$ for reasons of infrared safety, with $\ec$
the energy in the current hemisphere.  The choice of
$\mathcal{E}_{{\mathrm{lim}}}$ should be sensible (a not too small
fraction of $Q$) to avoid development of further significant
logarithms involving this quantity.\footnote{One could also replace
  the limit on $\ec$ with a limit on $\sum_{i\in \hc} |\vec{p}_i|$ ---
  this would reduce the sensitivity to non-perturbative effects
  associated with hadron masses \cite{MassEffects}. A procedure
  allowing one to avoid the cut altogether would be to normalise to the
  photon virtuality $Q$. In contrast to the thrust case
  \cite{DIStauz}, there would be no difference in the ensuing
  resummation.}

In the $1+1$ jet limit, the broadening is small and in the
perturbative expansion of the distribution each power of $\as$ can be
multiplied by up to two powers of $\ln B_{zE}$. This leads to a
very poorly convergent series and necessitates a calculation which
sums the dominant terms of the perturbative series at all orders --- a
resummation.

As we have already mentioned, techniques for the resummation of event
shapes in $\ee$ are well established
\cite{CTTW,CTW,JetRates,NewBroad,Cpar} and more recently there have
been extensions to DIS \cite{DIStauz}, to non-global variables
\cite{NonGlobal} and to multi-jet configurations
\cite{BDMZ,rhoLight}.

In general for a resummable variable $V$ one can show an
exponentiation of the large logarithms, which means that the suitably
normalised cross section for the variable to be smaller than some
value $V$ can be written in the following form:
\begin{equation}
  \label{eq:basicresummed}
  \sigma(V) = 
  \left(1 + \sum_n C_n \asb^n\right) e^{Lg_1(\as L) + g_2(\as
      L) + \cdots} 
     + D(V)\,,
\end{equation}
where $L = \ln 1/V$. The distribution is obtained by differentiating
this expression with respect to $V$.

The leading (or double) logs are those contained in $Lg_1(\as L)$ and
the next-to-leading (or single) logs are those contained in $g_2(\as
L)$. Further subleading sets of terms would be contained in functions
$\as^{n-2} g_n(\as L)$. There is also a remainder function $D(V)$
containing terms which go to zero for $V\to0$. For problems with
initial state hadrons (DIS, $p\bar{p}$) the $C_n$ and
$D$, as well as $g_2, g_3,\ldots$ are generally $x$
dependent and involve convolutions with parton distribution functions
(which in eq.~\eqref{eq:basicresummed} have not been explicitly
shown).

The summation of leading and NL logarithms is usually sufficient to
control the normalisation of the distribution down to the region of
the peak of the distribution, $\as L \sim 1$. The contribution from
further subleading terms, such as $\as g_3(\as L)$, will be of order
$\as$ in that region and so formally a `small' correction.

The variable considered here is however unusual in that if one follows
the standard procedure and keeps just the leading and next-to-leading
logs, calculated in section~\ref{sec:Derivation}, one obtains an
answer which \emph{diverges} for $\as L$ of order 
$1$. Such an occurrence is not unknown in problems 
involving initial state hadrons. 
In Ref.~\cite{CatNasTrenMan} it was observed for the case of 
Drell-Yan lepton pair 
production near threshold, that a factorial 
divergence occurs (shown to be unrelated to the expected renormalon behaviour) 
in the coefficients of the resummed formula 
 if one performs a naive resummation till NLL accuracy.
 
Additionally, problems closely related to those discussed here 
have been observed before in certain approaches
for calculating the transverse momentum distribution of
a Drell-Yan pair \cite{EllisStirling,RW},
but to our knowledge the DIS broadening represents the first time 
these difficulties have 
come up in the context of 
an event shape --- \ie a variable
with direct sensitivity to hadron emission.

Essentially the divergence stems from the fact that there is a 
dependence of the observable on 
the transverse momentum recoil. In
close analogy with the case of the Drell-Yan $p_t$ distribution, the
net vector sum of the recoil can go to zero in two ways: either by a
veto on emissions, or by vector sum of several emissions adding up to
zero \cite{DYPair}. The exponentiated form for the answer is suitable
for taking into 
account the first effect but not the second, and breaks down (with a
divergence) when the `easiest' (most likely) way of producing a low
transverse momentum recoil is through the second mechanism.

There are however important differences between the Drell-Yan case and
$B_{zE}$. The Drell-Yan transverse momentum is sensitive to emissions
exclusively through their recoil. The broadening however is sensitive
to emissions in the target hemisphere only through recoil, but to
emissions in the current hemisphere through both recoil and direct
`measurement' of the emissions. This means that standard solutions for
the Drell-Yan problem \cite{RW,DYPair} are not so easily applied to
$B_{zE}$.

Accordingly in section~\ref{sec:Solution} we develop a new technique
for supplementing the answer so as to avoid the divergence.
Essentially we redefine our accuracy criterion in terms of the
relative impact of a given contribution on the final answer rather
than in terms of a formal counting of logarithms. Technically this
requires the expansion of certain integrands to be carried out about a
point closer to their saddle-point than is needed in the standard
approach.  We show that after the application of this method remaining
uncertainties are pure (non-divergent) subleading terms.

For actual phenomenological applications, to be able to study the
event shape distribution over its full range, it is necessary to match
the resummed calculation to fixed-order results. There exist well
established techniques in $\ee$, however for a variety of technical
reasons they cannot be directly applied to the DIS case. So in
section~\ref{sec:MatchingMain} we examine the modifications that are
necessary as well as elaborating on a matching scheme proposed in
\cite{DIStauz} (here named $M$-matching) and introducing a new scheme
which we call $M_2$ matching.

A final element in the prediction of event shape distributions is
the non-perturbative correction. For most variables this is quite
straightforward, being a simple shift of the distribution over most of
its range \cite{KorchemskySterman,DokWebShift}. However for the
broadening, as is calculated in section~\ref{sec:NP}, in close analogy
with the $\ee$ case \cite{BroadPower}, the effect of the
non-perturbative correction is also to squeeze the distribution. One
interesting consequence of this is that the power correction to the
mean broadening acquires an $x$-dependent component, which has been
noticed experimentally by the ZEUS collaboration \cite{H1ZeusConf}.

Given all these ingredients we are therefore able for the first time,
in section~\ref{sec:Results}, to show a comparison of a resummed,
matched and power-corrected distribution to DIS event shape data
\cite{H1NewData}. In that section we also show how the resummed
results compare to the exact fixed order calculation and examine the
effect of the standard and improved resummations. Forthcoming work
\cite{DSDIS} will give a more detailed analysis of the data, for a
range of observables including the broadening.

In addition to the contents of the body of this paper described above,
there are several appendices containing details of the working of the
various sections. One appendix which we draw particular attention to
is appendix~\ref{sec:PDF}, which deals with the evolution of parton
distribution functions (PDF).  Though PDF evolution is not the subject
of this paper, it turns out that for the phenomenological
implementation of our formulae there are certain advantages (\eg
freedom in the
choice of the value of $\as$ for the evolution) to using one's own PDF
evolution rather than that embodied in standard PDF global fits such
as those from CTEQ, GRV or MRST \cite{CTEQ,GRV,MRST}.
Appendix~\ref{sec:PDF} discusses these advantages in detail, presents
the algorithms used, and also shows some discrepancies (though
fortunately in contexts of limited phenomenological importance) that
we have found in the evolutions embodied in the CTEQ5 and MRST99
distributions.

\section{Derivation}
\label{sec:Derivation}

Many aspects of the resummation of event shapes have become standard
in past years. Accordingly in this derivation we will shall be quite
concise, referring the reader to the literature
\cite{DIStauz,CTW,NewBroad} for a more detailed discussion of
certain subtleties.

\begin{figure}[htbp]
  \begin{center}
    \epsfig{file=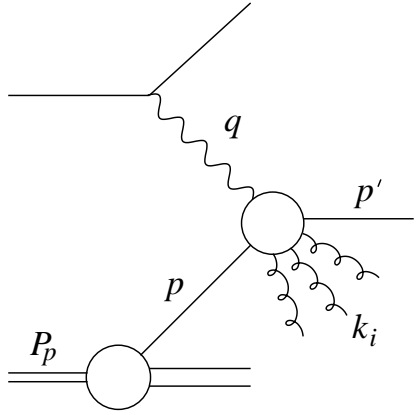,width=0.35\textwidth}
    \caption{Kinematics.}
    \label{fig:kin}
  \end{center}
\end{figure}
We start off by writing the momenta $k_i$ of radiated partons
(gluons and/or quark-antiquark pairs) in terms of Sudakov
(light-cone) variables as (Fig.~1)
\begin{equation}\label{eq:suddef}
 k_i\>=\>  \alpha_i P \>+\>\beta_i P' \>+\>k_{ti}\>,
\qquad\qquad \alpha_i\beta_i\>=\>\vec k_{ti}^2/Q^2\,,
\end{equation}
where $P$ and $P'$ are light-like vectors
along the incoming parton and current directions,
respectively, in the Breit frame of reference:
\begin{equation}
P\>=\>x P_p\>,\>\>
P'\>=\>x P_p+q\>,\>\> 
2(P\cdot P')\>=\>-q^2\>\equiv\>Q^2
\end{equation}
where $P_p$ is the incoming proton momentum and
$x=Q^2/2(P_p\cdot q)$ is the Bjorken variable. 
Thus in the Breit frame we can write
$P=\half Q(1,0,0,-1)$ and $P'=\half Q(1,0,0,1)$,
taking the current direction as the $z$-axis. Particles
in the {\em current hemisphere} $\hc$ have $\beta_i>\alpha_i$
while those in the proton {\em remnant hemisphere} $\hr$ have
$\alpha_i>\beta_i$.

In line with the procedure adopted in \cite{DIStauz}, we write the
following expression for the cross section (strictly speaking the
contribution to $F_2$ from an incoming quark of unit charge). It is
given in terms of $N$, the moment variable conjugate to Bjorken-$x$
for the case of emission of $m$ gluons in the remnant hemisphere and
$n$ in the current hemisphere, where all the gluons have $k_t \ll Q$,
\begin{equation}
  \label{eq:IndepEmission}
  \sigma_{mn} = 
  q_N(Q_0^2) \cdot \frac{1}{m!} \prod_{i}^{m} 
  \int_{Q_0^2}^{Q^2} \frac{d^2 \vkti}{\pi \kti^2} \frac{\as(\kti^2)
    \cf}{2\pi} 
   \rho_{R,i} \,\cdot\,
  \frac{1}{n!} \prod_{j}^{n} 
  \int^{Q^2} \frac{d^2 \vktj}{\pi\ktj^2} \frac{\as(\ktj^2) \cf}{2\pi}
   \rho_{C,j}\,.
\end{equation}
with the coupling defined in the Bremsstrahlung scheme \cite{CMW}
and
\begin{subequations}
  \label{eq:RrRt}
\begin{align}
  \label{eq:Rr}
  \rho_{R,i} &= \int z_i^N \; dz_i \,\frac{1 + z_i^2}{1-z_i}\,
  \Theta(Q\alpha_i- k_{ti}), \\
  \rho_{C,j} & = \int d\zbar_i \,\frac{1 + \zbar_i^2}{1-\zbar_i}
  \Theta(Q\beta_j-\ktj)\,.
\end{align}
\end{subequations}
Because of the collinear divergence on the incoming (proton) leg, we
have had to introduce a factorisation scale, $Q_0^2$. This scale
serves both for the parton distribution and as the lower limit on
transverse momentum for emissions in $\hr$.  The virtual corrections
are given by expressions which are similar except for the absence of
the factor $z^N_i$ in \eqref{eq:Rr}.

In the situation where there is an ordering  $k_{t1} > k_{t2} > \ldots
k_{ti}$, the $\alpha_i$ and $\beta_j$ can be written as follows
\begin{equation}
\alpha_i = \frac{1-z_i}{z_1 \ldots z_i}\,,\qquad\quad
\beta_j = \frac{1-\zbar_j}{\zbar_1 \ldots \zbar_j}\,. 
\label{eq:alphai}
\end{equation}
For differently ordered cases one should simply permute the indices
appropriately. In the limit of soft emissions the $z_i$ can all be
approximated by $1$. As is explained in \cite{DIStauz}, to our
accuracy it is actually possible to do this even when there are hard
collinear emissions. So the factor $\Theta(Q \alpha_i -\kti)$ can be
replaced by $\Theta(Q(1-z_i) -\kti)$ and analogously for $\Theta(Q
\beta_j -\ktj)$.

\subsection{$\boldsymbol{B_{zE}}$}

Here we work out the contribution to $F_2$ from events with a broadening
smaller than $B_{zE}$. We shall examine configurations consisting
solely of soft and/or collinear emitted gluons. In this limit the
difference between $ 2\sum_{i\in \hc} |\vec{p}_i|$ and $Q$ is small and given that the
broadening is also small we can replace $ 2\sum_{i\in \hc} |\vec{p}_i| \to Q$ introducing an
error of $\order{\kt^2/Q^2}$ which is negligible. In order for the
broadening to be smaller than $B_{zE}$, one then obtains the following
condition on the emitted momenta:
\begin{equation}
  \label{eq:BzETheta}
  \Theta \left(Q B_{zE} -   \sum_{\hc} k_{ti} - \left|\sum_{\hr,\hc}
    \vkti\right| \right) .
\end{equation}
The real emission part of the contribution to $F_2$ (for an incoming
quark of unit charge) is then given by
\begin{equation}
  \Sigma_N(B_{zE}) = \sum_{m,n}^\infty \sigma_{mn} \Theta \left(Q B_{zE} -
    \sum_{\hc} k_{ti} - \left|\sum_{\hr,\hc} 
    \vkti\right| \right) \,.
\end{equation}
In order to carry out a resummation we need not only $\sigma_{mn}$ in
factorised form, as given above, but also the $\Theta$ function. This
can be obtained with the aid of a couple of integral transforms:
\begin{equation}
\label{eq:AllTransforms}
 \Theta \to  \int \frac{d\nu }{2\pi i \nu } \frac{d^2\vb \, d^2\vpt}{4\pi^2}\,
  e^{\nu B} e^{-\nu p_t/Q} e^{i\vb \cdot \vpt}
   \prod_{i\in\hc} e^{-\nu \kti/Q} \prod_{i\in\hr,\hc} e^{i \vb \cdot
     \vkti}\,, 
\end{equation}
where we have explicitly introduced the current quark transverse
momentum $\vpt = - \sum_{i \in \hr,\hc} \vkti$. Carrying out the $p_t$
integration gives \cite{NewBroad}
\begin{equation}
  \Theta \to 
\int \frac{d\nu }{2\pi i \nu } \frac{Q^2 d^2\vb}{2\pi} \frac{\nu}{(\nu^2 + 
  b^2 Q^2)^{3/2}}\,
  e^{\nu B}
   \prod_{i\in\hc} e^{-\nu \kti/Q} \prod_{i\in\hr,\hc} e^{i \vb \cdot
     \vkti}\,. 
\end{equation}
We then define the double transform $\varsigma_N(\nu,b)$ of our
cross section:
\begin{equation}
  \Sigma_N(B) = \int \frac{d\nu}{2\pi i \nu} \frac{Q^2 d^2\vb}{2\pi}
  \varsigma_N(\nu,b).
\end{equation}
Writing the sums over $m$ and $n$ as exponentials one obtains the following 
all-orders expression for $\varsigma_N(\nu,b)$:
\begin{multline}
  \varsigma_N(\nu,b) = \exp\left\{ 
      \int_{Q_0^2}^{Q^2} \frac{d^2\vkt}{\pi k_t^2} 
      \frac{\as(k_t^2) \cf}{2\pi}
      \int_0^{1-\frac{k_t}{Q}} dz
      \frac{1+z^2}{1-z} \left(z^N e^{i\vb . \vkt} - 1\right) 
      \right.\\\left.
      +\int^{Q^2} \frac{d^2\vkt}{\pi k_t^2} 
            \frac{\as(k_t^2) \cf}{2\pi}
      \int_0^{1-\frac{k_t}{Q}} d\zbar
      \frac{1+\zbar^2}{1-\zbar} \left(e^{i\vb . \vkt - \nu k_t/Q} - 1\right)
    \right\} q_N(Q_0^2)\,.
\end{multline}
We have simplified the phase-space restrictions in
eq.~\eqref{eq:RrRt}, making the approximation that $z_1\ldots z_i
\simeq 1$, as explained in above. The terms $(-1)$ account for
the virtual corrections \cite{NewBroad}. We then write 
\begin{equation}
  q_N(Q_0^2) = \exp\left\{ -\int_{Q_0^2}^{Q^2} \frac{d^2\vkt}{\pi
      k_t^2} \frac{\as(k_t^2) \cf}{2\pi}
    \int_0^{1-\frac{k_t}{Q}} dz 
      \frac{1+z^2}{1-z} \left(z^N - 1\right) \right\} q_N(Q^2)\,,
\end{equation}
where, as elsewhere so far, we keep only the piece relating to gluon
emission from a quark. We have exploited the fact that whenever an
integrand is finite for $z\to1$ we are allowed to ignore the
(subleading) difference between $1-k_t/Q$ and $1$ in the upper limit
on $z$.  It is then convenient to rearrange the $N$-dependence so as to
isolate the $\gamma_{qq,N}$ anomalous dimension:
\begin{equation}
  \int_0^{1-\frac{k_t}{Q}} dz \frac{1+z^2}{1-z} \left[(z^N e^{i \vb .\vkt} -
    1) - (z^N - 1)\right] = (e^{i 
    \vb .\vkt} - 1) \left[\gamma_{qq,N} + 
    \int_0^{1-\frac{k_t}{Q}} dz \frac{1+z^2}{1-z} \right].
\end{equation}
Since the $N$-dependence is now completely separated from the
$1/(1-z)$ soft divergence it is straightforward to replace
$\gamma_{qq,N}$ with the full anomalous dimension matrix, $\bgamma_N$.
Accordingly our complete answer in Mellin transform space is:
\begin{equation}
  \varsigma_N(\nu,b) =  \bC_{0,N} \, e^{-\rc(\nu,b) - \rr(b)} \boldsymbol{q}_N(Q^2)\,,
\end{equation}
where $\bC_{0,N}$ is a matrix of zeroth order coefficient functions in $N$ space (see \cite{DIStauz}), $\boldsymbol{q}$ is a vector of parton distributions and 
one has the following expressions  for radiators $\rc$ and $\rr$:
\begin{subequations}
\begin{align}
  \rr(b) &= -\int \frac{dk_t^2}{k_t^2} \frac{\cf\as(k_t^2)}{2\pi}
  \left(J_0(b k_t) - 1\right)\left(\bgamma_N + \int_0^{1-\frac{k_t}{Q}} dz
    \frac{(1+z^2)}{1-z} \right),
  \\
  \rc(\nu, b) &= -\int \frac{dk_t^2}{k_t^2} \frac{\cf\as(k_t^2)}{2\pi}
  \left( J_0( b k_t)\, e^{-\nu k_t/Q} -1\right)\int_0^{1-\frac{k_t}{Q}}
  d\zbar \,\frac{1 + \zbar^2}{1-\zbar}\,.
\end{align}
\end{subequations}
To NLL accuracy these integrals can be evaluated by making the
following replacements \cite{NewBroad,RW}:
\begin{subequations}
\begin{align}
  \left( J_0( b k_t)\, e^{-\nu k_t} -1\right) &\to \Theta\left(k_t -
    \frac{2e^{-\gae}Q}{\nu +\sqrt{\nu^2 + Q^2b^2}}\right),
  \\
  \left( J_0( b k_t)\, -1\right) &\to \Theta\left(k_t - \frac{
      2e^{-\gae}}{b }\right).
\end{align}
\end{subequations}
Carrying out the integrations over  $\zbar$ and $z$ one obtains
\begin{subequations}
  \label{eq:RRCnub}
\begin{align}
  \label{eq:RCnub}
  \rc(\nu,b) &\simeq R\left(\frac{\nubar
      +\sqrt{\nubar^2 + Q^2\bbar^2}}{2}\right), \\
  \rr(b) &\simeq R_\gamma\left(\frac{\bbar Q }{2}\right),
\end{align}
\end{subequations}
where we have introduced $\nubar = \nu e^{\gae}$ and $\bbar = b
e^{\gae}$, and 
\begin{subequations}
\label{eq:RadiatorIntegrals}
\begin{align}
  R(u) &= \int_{Q^2/u^2}^{Q^2} \frac{dk_t^2}{k_t^2}
  \frac{\cf\as(k_t^2)}{2\pi} \left( 2\ln \frac{Q}{k_t} - \frac 32 \right),
  \\
  R_\gamma(u) &= \int_{Q^2/u^2}^{Q^2} \frac{dk_t^2}{k_t^2}
  \frac{\cf\as(k_t^2)}{2\pi} \left( 2\ln \frac{Q}{k_t} - \frac 32 +
    \bgamma_N\right).
\end{align}
\end{subequations}
Explicit expressions for $R(u)$ and $R_\gamma(u)$, to leading and
next-to-leading accuracy, are given in appendix~\ref{sec:radiators}.

Our answer in $\nu$ space is then given by an integral over $b$:
\begin{equation}
  \label{eq:sigmanu}
  \varsigma_N(\nu) =  \bC_{0,N} \int \bbar d\bbar \frac{\nubar}{(\nubar^2 +
    \bbar^2Q^2)^{3/2}} 
  e^{-R\left(\frac{\nubar
    +\sqrt{\nubar^2 + Q^2\bbar^2}}{2}\right) -
R_\gamma\left(\frac{Q\bbar}{2}\right)}   \boldsymbol{q}_N(Q^2) \,.
\end{equation}
To evaluate this integral, the procedure that has being adopted
previously \cite{NewBroad,DIStauz}, has been to expand the functions
$R(u)$, eqs.~\eqref{eq:RRCnub}, as
\begin{subequations}
  \label{eq:RExp}
  \begin{align}
    R\left(\frac{\nubar
      +\sqrt{\nubar^2 + Q^2\bbar^2}}{2}\right) &=
  R(2\nubar) + R' \ln \frac{1 +\sqrt{1 + y^2}}{4}  
  + \order{R'' \ln^2 y}
  + \order{R_2' \ln y}\,,
  \\
  \label{eq:RgammaExp}
    R_\gamma\left(\frac{\bbar Q}{2}\right) &=
  R_\gamma(\nubar) 
  + R_\gamma' \ln \frac{y}2
  + \order{R_\gamma'' \ln^2 y}
  + \order{R_{\gamma2}' \ln y}\,,
  \end{align}
\end{subequations}
where we have introduced $y =\bbar Q/\nubar$. We have defined $R'$ as
been the pure single-logarithmic piece of $\nu \partial_\nu R(\nu)$;
accordingly $R'_\gamma = R'$. The part of $\nu \partial_\nu R(\nu)$
containing terms $\as^n \ln^{n-1} \nu$ is referred to as $R_2'$ (and
analogously for $R_{\gamma}$). In what follows immediately below it
can be neglected because it leads only to NNLL corrections. The same
criterion also allows us to throw away the terms containing $R''$.

So we can now write our expression for the $\nu$-space resummed cross
section:
\begin{equation}
  \label{eq:BadResult}
  \varsigma_N(\nu) =    \bC_{0,N} \, e^{-R_\gamma(\nubar) -
    R(2\nubar)} \Lambda(2R') \,\boldsymbol{q}_N(Q^2)\,,
\end{equation}
with
\begin{equation}  
  \label{eq:Lambda}
\Lambda(2R') = \int_0^\infty
  \frac{y\,dy}{(1 + y^2)^{3/2}}\, 
  \left(\frac{1 +\sqrt{1 + y^2}}{4}\right)^{-R'}
  \left(\frac{y}{2}\right)^{-R'}\,.
\end{equation}
We are not aware of a closed form for $\Lambda$.  Its
expansion for small $R'$ is
\begin{equation}
  \label{eq:LambdaExpnsn}
  \Lambda(2R') = 1 + \frac{7\pi^2 - 36\ln^2 2}{96}(2R')^2\, + \order{{R'}^3}.
\end{equation}

The final step of the calculation is to perform the inverse Mellin
transform with respect to $\nu$:
\begin{equation}
  \Sigma_N(B) = \int \frac{d\nu}{2\pi i \nu}\, e^{\nu B}  \varsigma_N(\nu)\,.
\end{equation}
It can be shown in a variety of ways (see for example \cite{NewBroad})
that this gives
\begin{equation}
  \label{eq:invMellin}
  \Sigma_N(B) \simeq \frac{1}{\Gamma(1 - \nu 
    \partial_\nu \ln \varsigma_N(\nu)|_{\nu=1/B})}
  \varsigma_N\left(\frac{1}{B}\right) .
\end{equation}
Accordingly after throwing away all NNLL (and yet higher order) terms
we obtain
\begin{equation}  
\label{eq:StandardResult}
  \Sigma_N(B) =   \bC_{0,N} \, \frac{\Lambda(2R')}{\Gamma(1 + 2 R')}
  e^{-R(1/B)-R_\gamma(1/B) - R' (\ln 2 + 2\gamma_E) }
  \boldsymbol{q}_N(Q^2)
\,.
\end{equation}
Its fixed order expansion is given in terms of the coefficients
$G_{nm}$ of $(\as/2\pi)^n \ln^m 1/B$ in the exponent, which are listed
in table~\ref{tab:Gnm}. Note that in contrast with the convention
adopted in \cite{DIStauz} the change in scale of the parton
distribution has not been written explicitly, but rather left
implicit through the action of $R_\gamma$ on $\mathbf{q}_N$. Accordingly the
$G_{mm}$ are not pure numbers but rather operators in flavour space.

We note that the answer has been checked against an analytical
first-order calculation of the dominant terms at small $B$, given in
appendix~\ref{sec:FORes}. It has also been tested (strictly the form
\eqref{eq:xResum}, which includes the $\order{\as}$ constant term)
against fixed order results from DISASTER++ \cite{Disaster} and we
find good agreement for all terms that are intended to be under
control, namely $\asb L^n$ with $0\le n\le 2$ and $\asb^2 L^n$ with
$2\le n\le 4$.

Finally, we observe that while we have chosen to resum logarithms of
$1/B$, we could just as easily considered a resummation of say
logarithms of $1/XB$. With an appropriate change of the $g_2$
function, given in appendix~\ref{sec:Rescaling} one then obtains an
equivalent answer to NLL order, but with different effective
subleading dependence on $B$. Such a rescaling of the argument of the
logarithm is therefore a way of testing the sensitivity of the answer
to uncontrolled subleading effects.

\begin{table}[htbp]
\begin{center}
\renewcommand{\arraystretch}{1.8}
\begin{tabular}{|c|l|}\hline
  $G_{12}$ & $-4\CF$ \\\hline
  $G_{11}$ & $(6 - 4\ln 2)\CF - 2\bgamma_N$ \\\hline
  $G_{23}$ & $-\frac{32\pi\beta_0}{3}\CF$ \\\hline
  $G_{22}$ & $\left((12-16\ln 2)\pi\beta_0 - 4K -
    \left(\frac{2\pi^2}{3} + 24\ln^2 2\right)\CF \right)\CF -
            4\pi\beta_0\bgamma_N$\\
 \hline
\end{tabular}
\caption{Fixed order coefficients of the resummation for $B_{zE}$, at
  order $\as$ and $\as^2$.}
\label{tab:Gnm}
\end{center}
\end{table}


\subsection{Problems}
\label{sec:Problems}

Though eq.~\eqref{eq:StandardResult} is correct to NLL order, it turns
out that even in the region where the expansion is formally valid, $R'
\sim 1$, there are problems.  For $R' \ge 2$ the integral
\eqref{eq:Lambda} diverges in the $y\to0$ region. One can examine in more
detail what is happening by including the $R''$ term of our expansion
around $ bQ = \nu$ in 
eq.~\eqref{eq:RgammaExp}. One finds that for $R'$ close to $2$, the
modified form of the integral \eqref{eq:Lambda}, gives
\begin{equation}
 \Lambda = \frac{1}{2-R'} + \frac{R''}{(2-R')^3} + \cdots
\end{equation}
\ie the formally subleading $R''$ term (as well as yet higher-order
terms of the expansion) becomes enhanced and can no longer be
neglected.

The breakdown of our expansion around $b = \nu/Q$ is associated with
two physical facts: firstly, half of the double logarithmic contribution comes
solely from recoil of the current quark with respect to emissions in
$\hr$; secondly the recoil transverse momentum of the current quark can be
zero even if it has radiated gluons --- it suffices that the vector
sum of the emitted transverse momenta be zero. 

It has been known for some time \cite{DYPair} that there are two competing
mechanisms for obtaining a small recoil transverse momentum. One can
restrict the transverse momentum of all emissions in $\hr$ --- locally
the probability of getting a small transverse momentum from this
mechanism scales as $p_t^{R'}$ (there is another factor $p_t^{R'}$
coming from a restriction on emissions in $\hc$, however this factor
persists independently of any discussion of recoil because it is also
generated by 
directly observed gluons). Alternatively one can have a small recoil
transverse 
momentum due to the cancellation of larger emitted transverse momenta,
and the corresponding probability scales as $p_t^{2}$.

In the region where $R'< 2$ the easiest (least suppressed) way of
restricting the $p_t$ is to restrict emitted transverse momenta.
Accordingly one sees a probability proportional to a Sudakov form
factor. However for $R' > 2$ this Sudakov form factor associated with
$\hr$ emissions gets frozen (at its $R'=2$ value) and the alternative
mechanism of suppression takes over.

The divergence that we see is associated with this transition.  It
arises because we are trying to use a formula,
eq.~\eqref{eq:BadResult}, with the same double logarithmic Sudakov
structure above and below $R'=2$, and a single logarithmic factor
$\Lambda(2R')$ intended to account for the effects of multiple
emission. However there is no way for a single logarithmic function to
cancel the effect of a double-logarithmic Sudakov form factor and
bring about the ``freeze-out'' discussed above --- so at the point
where this is supposed to happen, the simple approach of
eqs.~(\ref{eq:RExp}-\ref{eq:Lambda}) breaks down, giving a
divergence.

Mathematically, what this corresponds to is that for very small $B$
the integral \eqref{eq:sigmanu} is dominated by values of $b Q \ll
\nu$ (related to individual emitted transverse momenta in $\hr$ being
much larger than $B Q$), so that an expansion around $bQ \sim \nu$ is
bound to fail.

Related problems have been seen before, in certain approaches for
the calculation of the transverse 
momentum distribution of 
a Drell-Yan pair \cite{EllisStirling,RW}. They turn out to be a general
feature of observables for which the contributions from different
emissions can cancel. Other examples of such variables are the
oblateness (the difference between the thrust major and minor)
\cite{NumSum} and the
difference between jet masses in $\ee$. Strictly speaking even the
thrust $\tau_{zE}$, 
resummed in \cite{DIStauz}, suffers from this problem --- however
there, for actual values of $\beta_0$, the divergence turns out to
be to the left of the Landau pole and so can be ignored.

\section{Beyond the divergence}
\label{sec:Solution}

For most phenomenological purposes it turns out that the divergence at
$R'=2$ does not 
cause any practical problems. This is because it is considerably to
the left of the maximum of the distribution ($R'=1/2$), in a region
where the distribution is strongly suppressed by the Sudakov form
factor, and where there are in any case large uncontrolled
non-perturbative corrections.

However one can envisage cases where the divergence may cause problems
(for example when using a non-perturbative shape function so as to
extend the distribution down to zero $B$
\cite{KorchemskySterman,GardiRathsman}), especially at low $Q$ values,
where it is more pronounced.  Furthermore it is in a region which
should formally be under control. So we feel that it is worth
dedicating some effort to improving the answer in this region.
Additionally the techniques that we develop may be of use for other
variables where similar divergences occur closer to the
phenomenologically relevant region.

Before entering into the details of the method it is perhaps worth
commenting on criteria for including subleading terms. In the standard
approach it is usual to keep the minimal set of terms --- \ie just
the leading and next-to-leading logs, and throw away anything which
contributes beyond this accuracy. This is analogous to the philosophy
in a fixed-order calculation, where one keeps only the orders one
knows and sets higher order terms to zero. Accordingly for example,
whenever we have an expression involving $\partial_L R$ we keep only the
dominant (single-log) terms $R'$, eq.~\eqref{eq:Rp}, because other
terms would lead to next-to-next-to-leading contributions.

For a `normal' observable, in the region $\as L \sim1$, the inclusion
of NLL terms is required to guarantee that the relative error on the
answer is of order $\as$ (associated with the NNL $\as g_3(\as L)$
term in the exponent). For the broadening a natural extension of
our accuracy criterion, is to keep not just a particular set of logarithms, 
but additionally all terms whose relative
contribution in the region $\as L\sim1$ is larger than $\order{\as}$,
even if they are formally N$^n$LL, with $n\ge2$. In general we will
try to take prescriptions which are as close to the original formulation
as possible.  We refer to these as `minimal' prescriptions.

\subsection{Improved resummation}

In the standard approach the divergence arises only when taking the
inverse Fourier and Mellin transforms of our answers. Accordingly up
to eq.~\eqref{eq:sigmanu} the resummation procedure remains
unmodified. It is in performing the inverse transforms that the method
needs to be improved. Roughly speaking our approach will involve
expanding $\cR_R(b)$ around a point $b_0$ close to the saddle point of
the integral \eqref{eq:sigmanu}, rather than around $b=Q/\nu$. It will
also be necessary to keep a larger number of terms in the expansion.

So we start off by finding the point $b_0$ of eq.~\eqref{eq:sigmanu}
around which to expand $\cR_R$. To within our accuracy, as is shown in
appendix~\ref{sec:accuracy}, it is necessary for $b_0$ to be close to
the saddle point of the integral, but it is allowed to differ from the true
saddle point by a factor of order $1$. This enables us to choose $b_0$
such that in the limit of small $R'$ we have $b_0 Q =\nu$, as in the
standard resummation.

A convenient way of doing this is to define $b_0 = y_0 \nu/Q$ through the
saddle point, $y_0$, of the following integral
\begin{equation}
  \Lambda_m = \frac{2}{\pi} \int \frac{dy}{y} \frac{1}{\cosh 2\ln y}
  \exp\left( -(L+\ln y)R_1(\as(L+\ln y))\right)\,,
\end{equation}
with $L = \ln \nu$. This corresponds to finding the solution of the
following equation, where we have defined $\ell = \ln y_0$:
\begin{equation}
  2 \tanh 2\ell + R'(\as(L+\ell)) = 0\,,
\end{equation}
or equivalently
\begin{equation}
  \label{eq:tanhlsoln}
  2 \tanh 2\ell +
  \frac{2\CF}{\pi}\frac{\as(L+\ell)}{1-2\as\beta_0(L+\ell)} = 0 \,.
\end{equation}
This prescription for choosing $b_0$ is minimal not just because for
small $R'$ it gives $b_0Q=\nu$, but also because all unnecessary
subleading terms have thrown away. The solution to
\eqref{eq:tanhlsoln} has the following expansion
\begin{equation}
  \label{eq:ellExpnsn}
  \ell = 
  -\CF \asb L - \CF L (4\pi 
  \beta_0 L - \CF)\asb^2 + \order{\as^3 L^3}\,,
\end{equation}
where we have defined $\asb = \as/2\pi$.
We also define
\begin{equation}
  \Lbar = L - \ell\,,\qquad 
  \Rpbar=R'(\as \Lbar)\,,\qquad 
  \Rppbar=R''(\as \Lbar)\,.
\end{equation}
We then expand $\cR_R(b)$ around $b_0 = \frac{\nu}{Q} e^{\ell}$,
\begin{multline}
  \label{eq:expansion}
  \cR_R(b) = R_\gamma(\nu e^{\ell}) + 
  \gae \Rpbar + 
  \left(\ln \frac{b
      Q}{2 \nu} - 
    \ell\right)(\Rpbar + \Rpbar_{\gamma 2} + \gae \Rppbar)\\
  + \frac12  \left(\ln \frac{b Q}{2 \nu} - \ell\right)^2 \Rppbar + 
  \frac16\left(\ln \frac{b Q}{2 \nu} - \ell\right)^3 \Rpppbar\,.
\end{multline}
Comparing to the expansion~\eqref{eq:RgammaExp}, the additional
$\Rppbar$ term on the second line is required in order to control the
answer to within a factor of $\order{1}$, while the
$\Rpbar_{2\gamma}$, $\gae \Rppbar$ and $\Rpppbar$ give contributions
of relative order $\sqrt{\as}$. This too is discussed in detail in
appendix~\ref{sec:accuracy}. For $\cR_C(\nu,b)$ one expands as before,
\begin{equation}
  \cR_C(\nu,b) = R(2\nubar) + \left(\ln \frac{1+\sqrt{1+ 
        b^2 Q^2/\nu^2}}{4}\right)R' +\cdots\,,
\end{equation}
since the series remains well-behaved in the limit $b\to0$. 

We then write our $\nu$-space answer as 
\begin{equation}
  \label{eq:NewExpansion}
  \varsigma_N(\nu) =  \bC_{0,N} \LambdaBar\,
  e^{-R(\nu e^{\ell}) - e^{\gae}\Rpbar - R(\nu) -(2+\gae)R'} 
     \boldsymbol{q}_N\left(Q^2 e^{-2\ell}/\nu^2 \right)\,,
\end{equation}
where
\begin{multline}
  \label{eq:LambdaBar}
  \LambdaBar = \frac{1}{1 - \frac{\pi^2}{3}\asb\CF} \int
  \frac{dy}{y} \frac{y^2}{(1+y^2)^{3/2}} 
       \left(\frac{1+\sqrt{1+y^2}}{4}\right)^{-R'} 
       e^{-(\ln \frac{y}{2} - \ell)\Rpbar 
           -\frac12(\ln \frac{y}{2} - \ell)^2\Rppbar}
       \times \\ \times
         \left[1 - \frac16\left(\ln \frac{y}{2} - \ell\right)^3 \Rpppbar
           -\left(\ln \frac{y}{2} - \ell\right)(\Rpbar_{\gamma2} +
           \gae \Rppbar)
           \right]
         \,.
\end{multline}
In the expansion of the exponent, terms whose relative contribution is
of order $\sqrt{\as}$ have been kept only at first order, whereas
other terms must be kept at all orders.  The factor
$1/(1-\pi^2\as\CF/3)$ is included so as to ensure that $\LambdaBar$ is
free of any $\order{\as}$ contribution in analogy with the standard
resummation. Writing its expansion as
\begin{equation}
   \LambdaBar = 1 + \sum_{m=2}^\infty\sum_{n=0}^m \LambdaBar_{mn} \asb^mL^n
\end{equation}
the order $\as^2$ terms are given by
\begin{subequations}
\label{eq:LambdaBarExpnsn}
\begin{align}
  \label{eq:LambdaBar20}
  \LambdaBar_{20} &= \frac{3}{10}\,{\CF}^{2}{\asb}^{2}{\pi }^{4}
  + \zeta(3)[6\bgamma_N +(12\gae- 9)\CF
   - 8\pi\beta_0]\CF\,,
  \\
  \LambdaBar_{21} &= \frac43\,\CF\,\left (-2\,{\pi }^{3} \beta_0 + 9\,
    \CF\,\zeta (3)+6\,\CF\,X_{12}\right )+ \nonumber\\
  \label{eq:LambdaBar21}
  &\qquad\qquad\qquad
  +\CF(2\bgamma_N +(4\gae-3)\CF)\left(-1 + \frac{7\pi^2}{6} - 2\ln^2
    2\right) \,,
  \\
  \label{eq:LambdaBar22}
  \LambdaBar_{22} &= \frac23\,\CF^{2}\left (7\,{\pi }^{2}-36\,\ln^2
    2-6 \right )\,,
\end{align}
\end{subequations}
with
\begin{equation}
  X_{12} = \int \frac{dy}{y} \frac{y^2}{(1+y^2)^{3/2}}
  \ln \left(\frac{1+\sqrt{1+y^2}}{4}\right) \ln^2\frac{y}{2}
  \simeq 1.945031318\,.
\end{equation}
We note that eqs.~\eqref{eq:LambdaBar} and \eqref{eq:LambdaBarExpnsn}
involve the anomalous dimension matrix $\bgamma_N$, through its presence in the
term $\Rpbar_{\gamma 2}$.

It is useful to study the behaviour of the various factors of
\eqref{eq:NewExpansion} in the two regimes, $R' < 2$ and $R' > 2$. For
small $L$, the series expansion of $\ell$, eq.~\eqref{eq:ellExpnsn},
shows that $\ell$ remains small compared to $L$. Indeed $R_\gamma(\nu
e^\ell)$ differs from $R_\gamma(\nu)$ by a term of order $\as^2 L^2$
(and higher) which is exactly compensated by the $\order{\as^2L^2}$
difference between $\LambdaBar$ and $\Lambda$, \cfr
eqs.~\eqref{eq:LambdaBar22} and \eqref{eq:LambdaExpnsn}.

If one increases $L$, then one finds that a transition takes place
around $R' \simeq 2 $. Beyond this point $\ell$ starts to vary much
more rapidly, going roughly as
\begin{equation}
  \ell \simeq -L +  \frac{\pi}{(\CF + 2\pi\beta_0)\as}+
  \order{1}\,,
  \qquad\quad (R' > 2)
\end{equation}
Accordingly $R_\gamma(\nu e^{\ell})$ stops varying for $R' > 2$; on the
other hand $\LambdaBar$ starts varying rapidly, going as $e^{-2L}$.

In section~\ref{sec:Problems} we mentioned the presence of two
competing mechanisms for the current quark to have a small transverse
momentum. The transition in the behaviour of $\ell$ that we have just
discussed corresponds precisely to the transition from the mechanism
of suppression of radiation (associated with a Sudakov form factor),
to that of arranging for the vector sum of the emitted momenta to be
small (the probability of which scales as $k_t^2$).

It is important to understand this transition, and in particular the
$L$ dependence of the various factors in \eqref{eq:NewExpansion} in
order to perform the inverse Mellin transform of our result. As in
section~\ref{sec:Derivation} we shall make use of
eq.~\eqref{eq:invMellin}. To calculate the argument of the
$\Gamma$-function we need to know which factors in
\eqref{eq:NewExpansion} vary rapidly. Since we aim to control our
answer to a relative accuracy of $\order{\as}$ this means that (logs
of) terms whose derivatives are of order $\as L$, or of order $1$,
must be kept, while terms for the which the derivative is of order
$\as$ can be neglected.

So whereas in section~\ref{sec:Derivation} this meant that we could
neglect $\Lambda$, now, since $\LambdaBar$ can vary rapidly it needs
to be taken into account in $\nu\partial_\nu \ln \varsigma_N(\nu)$.
Nevertheless, in line with the approach of neglecting non-essential
subleading terms, we shall not take the derivative of the whole
function $\ln \varsigma_N(\nu)$, but only of those terms which are
essential. One possibility would be to define 
\begin{equation}
  \label{eq:RpeffWithGamma} 
  R'_{\eff} = \frac{d}{dL} ( \Lbar R_1(\as \Lbar) - \ln \LambdaBar)
\end{equation}
This however presents some technical difficulties, because
$\LambdaBar$ involves the anomalous dimension matrix. These
difficulties could perhaps be surmounted, however a simpler solution
is to observe that while $\Rpbar_{\gamma 2}$ contributes pieces of
relative order $\sqrt{\as}$, they vary significantly only over a
region of $\Delta L \sim 1/\sqrt{\as}$ (see
appendix~\ref{sec:accuracy}).  Accordingly in $\partial_L \ln
\LambdaBar$ they can at most contribute an amount of order $\as$ and
so can be neglected. So to avoid the complications associated with the
anomalous dimension matrix, in \eqref{eq:RpeffWithGamma}  we could
replace $\Lambda$ with the following `simpler' quantity:
\begin{multline}
  \label{eq:LambdaBars}
  \LambdaBar_n = \frac{1}{1 - \frac{\pi^2}{3}\asb\CF} \int
  \frac{dy}{y} \frac{y^2}{(1+y^2)^{3/2}} 
       \left(\frac{1+\sqrt{1+y^2}}{4}\right)^{-R'} 
       e^{-(\ln \frac{y}{2} - \ell)\Rpbar 
           -\frac12(\ln \frac{y}{2} - \ell)^2\Rppbar}
       \times \\ \times
         \left[1 - \frac16\left(\ln \frac{y}{2} - \ell\right)^3 \Rpppbar
           -\left(\ln \frac{y}{2} - \ell\right)(\Rpbar_{2} +
           \gae \Rppbar)
           \right]
         \,.
\end{multline}
which differs from from $\LambdaBar$ only through the replacement of
$\Rpbar_{\gamma 2}$ by $\Rpbar_2$ in the second line. Correspondingly
its fixed order expansion differs from that of $\LambdaBar$ by the
absence of the $\bgamma_N$ terms in \eqref{eq:LambdaBar20} and
\eqref{eq:LambdaBar21}. The subscript $n$ indicates that this is a
non-minimal choice for $\LambdaBar$.

We can also make a more extreme choice, throwing away all terms which
do not contribute significantly to the derivative. One possibility
makes use of the same integrand as was used to determine the
saddle-point of $\ell$, and gives
\begin{equation}
  \label{eq:LambdaBarm}
  \LambdaBar_m = \frac{2}{\pi} \int \frac{dy}{y} \frac{1}{\cosh 2\ln y}
  \exp\left(-(\ln y - \ell)\Rpbar 
           -\frac12(\ln y - \ell)^2\Rppbar\right),
\end{equation}
where the subscript `$m$' stands for `minimal'.
The terms of its fixed order expansion that will be needed are
\begin{equation}
  \LambdaBar_{m21} = -\pi^3 \beta_0 \CF\,,\qquad\quad
  \LambdaBar_{m22} = \left(\frac{\pi^2}{2} - 4\right) \CF^2\,.
\end{equation}
Our final answer for the integrated broadening distribution will
therefore be given by
\begin{equation}
  \label{eq:FinalResult}
  \Sigma_N(B) = 
  \bC_{0,N} \frac{\LambdaBar}{\Gamma(1 + R' + R'_{\eff,X})}
  e^{-R(1/B)-R(1/\Bbar) - R' (\ln 2 + \gamma_E) - \Rpbar \gamma_E }
  \boldsymbol{q}_N(\Bbar^2Q^2) \,, 
\end{equation}
where $\Bbar = e^{-\ell} B$ and
\begin{equation}
  R'_{\eff,X} = \frac{d}{dL} \left( \Lbar R_1(\as \Lbar) - \ln
    \LambdaBar_X \right),
\end{equation}
with $X=s$ or $m$. Here, we have written the scale of the parton
distribution explicitly, to emphasise that it is now $\Bbar Q$ rather
than $BQ$.

For the purposes of matching to fixed order we need to know the
expansion of \eqref{eq:FinalResult} to $\order{\as^2}$.  One finds
that it differs from that of \eqref{eq:StandardResult} by the
following additional (subleading) terms:
\begin{equation}
\left[  \LambdaBar_{20} - \LambdaBar_{X21} \gae 
     + \left(\LambdaBar_{21} + (2\bgamma_N - 3\CF)\CF - (2\LambdaBar_{X22}
     + 4\CF^2)\gae\right) L \right] \asb^2\,.
\end{equation}


\section{Matching}
\label{sec:MatchingMain}

In order to extend the range of validity of the predictions for
event-shape distributions, various procedures have been developed in
$\ee$ \cite{CTTW} for supplementing resummed distributions with the information
from the first and second fixed order distributions. This is referred
to as \emph{matching}. 
It is not possible to merely carry over the matching 
schemes developed in $\ee$ to DIS without addressing certain technical 
complications that arise in the DIS context, which shall become evident here.
Additionally we propose new 
schemes that are suitable for the purposes of matching our resummations to fixed order estimates.
The discussion that follows will be kept fairly general since we intend to use 
the various schemes and ideas introduced here 
not just for the jet broadening but for other DIS variables 
that have been resummed 
thus far \cite{DIStauz,DSDIS}. For this reason we refer to all
distributions as being a function of a generic variable $V$ rather
than specifically of the broadening $B$.

We begin by examining the form of our resummed cross section. This 
has the following structure in moment ($N$) space
\begin{equation}
\label{eq:Nresum}
\sigma_{r,N}(Q^2,V) = \left [ {\bC}_{0,N} + 
\bar{\alpha}_s {\bC}_{1,N} \right ] \exp [L {\bg}_1(\alpha_s
L)+{\bg}_2(\alpha_s L)]\, \boldsymbol{q}_N(Q^2),
\end{equation}
where the subscript $N$ denotes Mellin transformed quantities and 
we have introduced flavour vectors and matrices in $2 n_f + 1 $
dimensions for the constants,  
the parton distributions (see \cite{DSDIS} for details) and the
functions $g_1$ and $g_2$. The operator structure for ${\bg}_1$ is
trivial (diagonal matrices with  the scalar function $g_1$, computed
as in appendix \ref{sec:radiators}  
for the broadening case,
for the quark entries and zero for the gluon entry)  
and only needed for dimensional consistency. The operator structure of
$\bg_{2}$ is non-trivial due to the presence of the anomalous
dimension operator. 
Note that this result is for the un-normalised cross section. 
Since a normalisation in $N$ space does not correspond to 
a normalised $x$ space result, we only normalise 
our result {\it{after}} translating to $x$ space.

The normalised resummed result in $x$ space then reads as 
\begin{equation}
\label{eq:xResum}
\sigma_r(x,Q^2,V) = \Sigma(x,Q^2,V) + \bar{\alpha}_s \exp[Lg_1+g_2^*]{\bC}_1 \otimes \frac{{\boldsymbol{q}}(x,V^nQ^2)}{q(x,Q^2)},
\end{equation}
with the {\it{form factor}} $\Sigma$ defined by 
\begin{equation}
\label{eq:formfac}
 \Sigma (x,Q^2,V) = \exp[L g_1+g_2^*] \frac{q(x,V^n Q^2)}{q(x,Q^2)}\,,
\end{equation}
where we have introduced the singlet distribution 
\begin{equation}
  \label{eq:qdef}
  q(x,Q^2) = \sum_{j=u,d,s,\ldots} e_j^2 \,[q_j(x)+\bar{q}_j(x)]\,,
\end{equation}
and used it to normalise the $x$ space result.
Note that the piece of $\bg_2$ corresponding to DGLAP evolution 
(that involving the anomalous dimension matrix) has been used 
to change the scale of the parton distributions 
to $V^n Q^2$ leaving behind the $g_2^*$ function. 
All quantities in \eqref{eq:formfac} 
are now scalars rather than operators 
since in writing the above we have multiplied out the matrices involved. 
The result for ${\bC}_1$ is available in appendix \ref{sec:FORes}.
The index $n$ can have the values 0 (for variables like the jet mass, C parameter 
or thrust with respect to thrust axis), 1 (for the thrust with respect to the photon axis) 
or 2 (for the broadening).

A point that we wish to draw attention to is that although the form
factor contains a parton distribution evaluated at scale $V^n Q^2$, we
can
ignore this change of scale in 
the second ($\order{\alpha_s}$) term of eq.~\eqref{eq:xResum} (and use $Q^2$ 
for the scale of the parton distribution) 
since it leads to subleading terms starting at the 
$\alpha_s^2 \ln 1/V$ level. 
However such a term will of course be relevant in the matching to NLO. 
Hence if one chooses not to keep the scale as $V^n Q^2$ in the convolution 
involving $C_1$ 
we will have to modify the matching piece accordingly.

Now we are ready to match the resummed result to 
the fixed order result returned by the NLO DIS Monte Carlo programs. 
Let us denote these exact results by $\sigma_e^{(1)}$ and $\sigma_e^{(2)}$ 
for the $\alpha_s$ and $\alpha_s^2$ Monte Carlo estimates. These are the 
result of a convolution with a specified structure function and so are
returned in $x$ space, and taken normalised to $q(x,Q^2)$.   
In order to perform the matching we essentially have to add the Monte Carlo 
and resummed results and remove the pieces which would be double counted.
These pieces would be the $\order\as$ and $\order{\as^2}$ terms of the 
resummed result  $\sigma_r^{(1)}$ and $\sigma_r^{(2)}$, which can be obtained 
by expanding eq.~\eqref{eq:xResum}.

However note that the matched resummed 
cross section has to satisfy 
certain requirements. 
The most important property is that in the $V \to 0$ limit the cross section 
must vanish on physical grounds \cite{CTTW}.
Accordingly the following matching formula is invalid
\begin{equation}
  \label{eq:NaiveMatching}
  \sigma(V) = \sigma_r + \asb \left(\sigma_e^{(1)} -
    \sigma_r^{(1)}\right) + \asb^2 \left(\sigma_e^{(2)} -
    \sigma_r^{(2)}\right)\,,
\end{equation}
because for $V \to 0$, the factor
$(\sigma_e^{(2)} - \sigma_r^{(2)})$ does not vanish, but
rather grows as $\ln 1/V$.

In $\ee$ the two main matching procedures are $R$ and $\ln R$
matching ($R$ in the original papers is the equivalent of $\sigma$
here). In $R$ matching one determines (from the fixed-order
distribution) the $G_{21}$ and $C_2$ coefficients for the
distribution, and defines an improved resummation formula
\begin{equation}
  \sigma_R = (1 + C_1 \asb + C_2 \asb) e^{L g_1(\as L) + g_2(\as L) +
     G_{21} \asb^2 L}\,.
\end{equation}
Then the equivalent of eq.~\eqref{eq:NaiveMatching} with $\sigma_r$
replaced by $\sigma_R$ does indeed vanish in the limit $V\to0$.  This
procedure is feasible in $\ee$ because $C_2$ and $G_{21}$ are simple
constants, and they can be evaluated by subtracting $\sigma_r^{(2)}$
from $\sigma_e^{(2)}$ in the very small $V$ limit. However in DIS
$C_2$ and $G_{21}$ both have $x$-dependence, and it is simply not
feasible to extract numerically them with their $x$-dependence. One
might think of extracting them (individually for each $x,Q^2$ point)
once the convolution has been done with the structure functions.
However experience in $\ee$ shows that one needs to go to very low
values of $V$, with vast statistics, in order to reliably extract
these quantities. In DIS with DISASTER++, low values of $V$ are often
not accessible because of cutoff effects.  Furthermore the Monte Carlo
statistical errors tend to be an order of magnitude larger than for a
similar number of $\ee$ events, and a similar number of events in DIS
with DISASTER++ takes an order of magnitude more time to generate.

The $\ln R$ matching approach is more easily extended to DIS. The
philosophy of $\ln R$ matching is to carry out the matching in the
logarithm of the cross section rather than in the cross section
itself:
\begin{equation}
\label{eq:lnRprelim}
\sigma_V = \exp \left[ \ln \sigma_r + 
  \asb \left( (\ln \sigma_e)^{(1)} - (\ln \sigma_r)^{(1)} \right)
  +
  \asb^2 \left( (\ln \sigma_e)^{(2)} - (\ln \sigma_r)^{(2)} \right)
\right]\,,
\end{equation}
where $(\ln \sigma_e)^{(n)}$ is the $\order{\as^n}$ part of $\ln
\sigma_e$. 

Strictly, taking the logarithm of the cross section is a
delicate operation because of the operator structure, which enters at
NLL level, in particular in the coefficients $G_{nn}$,
and one cannot for example use 
$\ln R$ matching exactly in the form prescribed in \cite{CTTW} .
However using 
\begin{equation}
\ln \sigma_{e,r} = \ln(1+\asb \sigma^{(1)}_{e,r}+\asb^2 \sigma^{(2)}_{e,r})   
\end{equation}
and expanding the logarithm in each case in powers of $\as$ one can alternatively write
\begin{equation}
  \label{eq:lnR}
  \sigma(V) = \sigma_r \, e^{\asb\left(\sigma^{(1)}_e - \sigma^{(1)}_r\right)
     + \asb^2\left(\sigma^{(2)}_e - \sigma^{(2)}_r 
        - \frac12(\sigma^{(1)}_e)^2 + \half(\sigma^{(1)}_r)^2
      \right)}\,,
\end{equation}
which still retains the correct $\order{\as^2}$ expansion as well as
the LL and NLL terms. Hence the above form is the one that should be used 
for $\ln R$ type matching in DIS.
Note that further subleading contributions are of course
not exponentiated as operators. But since they are beyond our
accuracy, we are entitled to mistreat them, 
as long they do not lead to some particularly 
pathological behaviour.

Another matching scheme we considered in \cite{DIStauz} 
was the following which we
shall now call multiplicative, or $M$-matching, 
\begin{multline}
  \sigma(V) = \sigma_r + \left[\asb \left(\sigma_e^{(1)} -
    \sigma_r^{(1)}\right) + \asb^2 \left(\sigma_e^{(2)} -
    \sigma_r^{(2)}\right) - 
  \right. \\ \left.
  \asb^2 \left(\sigma_e^{(1)}-
  \sigma_r^{(1)}\right)(L^2 G_{12} + L {\cG}_{11})
\right] \Sigma(x,Q^2,V)\,,
\end{multline}
where the presence of the form factor, $\Sigma$,  ensures that the whole cross
section does go to zero for $V \to 0$ and we have used ${\cG}_{nm}$ 
for the $x$ space versions of the resummation coefficients listed in Table 1.   
Note that the ${\cG}_{11}$ in the above result involves matrix products
and convolutions in $x$ space. 
For example for the jet broadening one gets from Table 1, by inspection 
\begin{equation}
\label{eq:G11proj}
{\cG}_{11}(x) = (6-4 \ln 2)\,C_F  - 2 \boldsymbol{C}_0 \otimes 
\boldsymbol{P} \otimes \frac{\boldsymbol{q}}{q},  
\end{equation}
 with $\bP$ being a matrix of leading order splitting functions.
The $x$ space versions of the $G_{n,n+1}$ coefficients 
are the same as the $N$ space numbers mentioned in Table 1 and we do
not distinguish them notationally.

We can also define $M_2$ matching:
\begin{equation}
  \sigma(V) = \sigma_r + \asb \left(\sigma_e^{(1)} -
    \sigma_r^{(1)}\right) + \asb^2 \left(\sigma_e^{(2)} -
    \sigma_r^{(2)}\right) \Sigma(x,Q^2,V) \,,
\end{equation}
which exploits the fact that the $\order\as$ term does vanish in the
$V\to 0$ limit. (Assuming of course that one has the correct $C_1$).
This has some similarity to $R$ matching in that the $\order{\as}$
piece of the remainder is not suppressed by a form factor.

If there is also resummation off a gluon\footnote{We encounter this
  situation in DIS observables such as current jet-mass, C parameter
  and the thrust defined with respect to the actual thrust axis
  \cite{DSDIS} as well as for light jet masses and narrow jet
  broadenings in $\ee $ \cite{rhoLight,DSDIS}}
then we need to modify the
matching somewhat. Strictly one would want some way in the fixed-order
contribution of separating out contributions associated with the
presence of just two gluons in the current hemisphere (or one gluon
plus virtual corrections). However with the existing tools, DISENT and
DISASTER++, this is not possible. Accordingly we arbitrarily choose to
attribute the entire difference between exact fixed order and the
expanded resummation to the part of the resummation that is off a
quark leg. The formulae for $M$ and $M_2$ matching remained unchanged, 
while that for $ \ln R $ matching becomes (where $\sigma_{rq}$, $\sigma_{rg}$ 
refer to the resummations off the current quark and gluon respectively)
\begin{equation}
\sigma_V = \sigma_{rg} + \exp \left[ \ln \sigma_{rq} + 
  \asb \left( (\ln \sigma_{eq})^{(1)} - (\ln \sigma_{rq})^{(1)} \right)
  +
  \asb^2 \left( (\ln \sigma_{eq})^{(2)} - (\ln \sigma_{rq})^{(2)} \right)
\right]
\end{equation}
where we have defined $\sigma_{eq} = \sigma_e - \sigma_{rg}$. After
explicitly taking the logarithms we obtain
\begin{equation}
  \label{eq:lnRWithGluon}
  \sigma(V) = \sigma_{rg} + 
  \sigma_{rq} \, e^{\asb\left(\sigma^{(1)}_e - \sigma^{(1)}_r\right)
     + \asb^2\left(\sigma^{(2)}_e - \sigma^{(2)}_r 
       -\frac12\left(\sigma^{(1)}_e - \sigma^{(1)}_r\right)
       \left(\sigma^{(1)}_e + \sigma^{(1)}_r - 2\sigma^{(1)}_{rg}\right)
      \right)}\,.
\end{equation}

There are some other 
important requirements of the final matched result. 
One is that at the upper limit of the 
distribution $V_{\mathrm{max}}$ the integrated cross section must go
to its {\it{exact}} 
upper limit without any additional leftover tems $\order{\as^3}$. Where the 
fixed-order differential distribution goes to zero at the upper limit 
one must ensure that the matched-resummed one does the same. 
In order to obtain 
these properties, 
one should use \emph{modified} matching formulae. The details can
be found in appendix~\ref{sec:ModMatch}.

\section{Non-perturbative effects}
\label{sec:NP}

For most event-shape observables, in the Born limit the principal
consequence of non-perturbative (NP) corrections is a uniform shift of
the distribution by an amount of order $\Lambda_{\mathrm{QCD}}/Q$
\cite{KorchemskySterman,DokWebShift}.\footnote{More precisely it is to
  convolute the perturbative distribution with a non-perturbative
  shape function of width $\Lambda_{\mathrm{QCD}}/Q$ \cite{KorchemskySterman},
  however as long as this width is much smaller than the width of the
  PT distribution, this effectively reduces to a shift.}  This is
because the effect of low-momentum radiation on the observable is
independent of the configuration of the hard momenta in the event.

For the broadening the situation is more complicated because the
effect of low-momentum emissions depends critically on the
configuration of the hard momenta. There are actually several effects.
One is that it is possible to neglect the recoil from the
non-perturbative emission because after azimuthal averaging it is
washed out relative to the recoil from perturbative radiation.
Accordingly the only NP correction to the broadening comes directly
from the transverse momentum of low-momentum particles emitted into
the current hemisphere. (This is similar to the effect which reduces
the naively calculated power correction to the heavy-jet mass by a
factor of two \cite{AkhouryZakharov,Milan2}).

But there is a second very important effect: soft gluons are emitted
uniformly in rapidity with respect to the quark axis. However
transverse momentum is measured with respect to the $z$ axis. When
gluons are emitted at very small angles to the quark axis the
contribution from their transverse momenta is entirely cancelled by a
longitudinal recoil of the quark. So only when gluons are emitted with
an angle larger than $\theta_{zq}$, the angle between the quark and
$z$-axes, do they contribute to the NP correction.

This has been discussed in detail for the $\ee$ broadening in
\cite{BroadPower}, and the 
techniques developed there can be carried through almost in their
entirety. Accordingly we shall only outline the steps, following very
closely the working of section 3.2 of \cite{BroadPower}, and
illustrating the relatively small differences.

\subsection{Power correction to the distribution}

After integrating over the rapidity of NP emissions, the NP
contribution to the broadening can be written as
\begin{equation}
  \cP \left( \ln \frac{Q}{p_t} + \eta_0\right) \,,\qquad \eta_0 \simeq
  -0.6137056\,,
\end{equation}
where $p_t$ is the transverse momentum (with respect to the $z$-axis)
of the current quark, and $\cP$ governs the overall magnitude of the
power correction \cite{DokWeb}:
\begin{equation}
  \label{eq:cPfin}
 \cP\> \equiv\>  \frac{4C_F}{\pi^2}\cM \frac{\mu_I}{Q}
\left\{ \alpha_0(\mu_I)- \as
  -\beta_0\frac{\as^2}{2\pi}\left(\ln\frac{Q}{\mu_I} 
+\frac{K}{\be_0}+1\right) \right\}\>,  \quad \as\equiv \al_{\MSbar}(Q)\>,
\end{equation}
with $\alpha_0(\mu_I)$ a non-perturbative parameter (corresponding
to the first moment of $\as$ up to some infrared scale $\mu_I$) which
is postulated to be observable and process independent. The `Milan'
factor $\cM$ accounts for the non-inclusiveness of the observable
\cite{Milan1,Milan2,MilanDIS}.

The effect of the non-perturbative contribution on the distribution
can be determined by replacing 
\begin{equation}
  e^{\nu B} \to e^{\nu B - \nu \cP\left(\ln \frac{Q}{p_t} +
      \eta_0\right)} 
\end{equation}
in eq.~\eqref{eq:AllTransforms}. Since $\cP$ is a small quantity, we
are allowed to expand the exponential and we write
\begin{equation}
  \varsigma_N(\nu) = \varsigma^{(PT)}_N(\nu) + \nu \cP f_N(\nu) + \order{\cP^2}\,,
\end{equation}
where the non-perturbative information function $f_N(\nu)$ is given by
\begin{multline}
  f_N(\nu) = 
  \bC_{0,N} \int_0^\infty   \frac{y\,dy}{(1 + y^2)^{3/2}}\, e^{-\cR_R(y\nu/Q ) -
    \cR_C(\nu,y\nu/Q)}
  \\
  \left(2 - \gae - \eta_0 + \ln\frac{1+\sqrt{1+y^2}}{2\nu(1+y^2)} -
    \sqrt{1+y^2}\right) \boldsymbol{q}_N(Q^2)
  \,.
\end{multline}
We make use of the results~\eqref{eq:RRCnub}, expand in powers of
$R'$, and introduce the differential representation of
\cite{BroadPower}, to obtain 
\begin{multline}
  f_N(\nu) = 
  \bC_{0,N} e^{-R(e^{\gae}e^{-\partial_a}) - R(2 e^{\gae}e^{-\partial_a})}
  \int_0^\infty   \frac{y\,dy}{(1 + y^2)^{3/2}}\, 
  \left(\frac{1 +\sqrt{1 + y^2}}{4}\right)^{-a/2}\\
  \cdot\left.
  \left(\frac{y}{2}\right)^{-a/2}
  \left(2 - \gae - \eta_0 + \ln\frac{1+\sqrt{1+y^2}}{2(1+y^2)} +\partial_a -
    \sqrt{1+y^2}\right) \boldsymbol{q}_N(Q^2)\, \nu^{-a}\right|_{a=0}
  \,.
\end{multline}
Evaluating the integral gives
\begin{multline}
  f_N(\nu) = \bC_{0,N} \, e^{-R(e^{\gae}e^{-\partial_a}) - R(2
    e^{\gae}e^{-\partial_a})}
  \Lambda(a) \\ \left.
  \left(2 - \gae - \eta_0 + \partial_a
   + \cI(a) + \cJ(a)  - \frac{8^{a/2}}{a}\right) \boldsymbol{q}_N(Q^2)\, \nu^{-a}
   \right|_{a=0}, 
\end{multline}
where
\begin{equation}
  \cI(a) = \frac{1}{\Lambda(a)}\int_0^\infty \frac{y\,dy}{(1 + y^2)^{3/2}}\, 
  \left(\frac{1 +\sqrt{1 + y^2}}{4}\right)^{-a/2}
  \left(\frac{y}{2}\right)^{-a/2}
  \ln\frac{1+\sqrt{1+y^2}}{2(1+y^2)}\,,
\end{equation}
and
\begin{multline}
  \cJ(a)  = -\int_0^\infty \left[\frac{y}{\Lambda(a)(1 + y^2)^{3/2}}\, 
  \left(\frac{1 +\sqrt{1 + y^2}}{4}\right)^{-a/2}
  \left(\frac{y}{2}\right)^{-a/2} 
  \sqrt{1+y^2} \right. \\ \left. - \frac{8^{a/2}}{(1+y)^{1+a}}\right]dy\,.
\end{multline}
Note the presence of the factor $8^{a/2}$ in the subtraction term,
required for a proper regularisation of the integral in the $a\to0$
limit. 

After carrying the $\nu$ integration and accounting for the effect of
the $\partial_a$ derivative, we obtain
\begin{equation}
  \Sigma_N(B) = \bC_{0,N} e^{-R(e^{\gae}e^{-\partial_a}) - R(2
    e^{\gae}e^{-\partial_a})}
  \left. \frac{\Lambda(a)}{\Gamma(1+a)} \left[ 1 + \frac{a}{B} \delta
    B(B,a)\right] \boldsymbol{q}_N(Q^2)\,  B^a \right|_{a=0}\!\!\!\!\!,
\end{equation}
with
\begin{equation}
  \label{eq:fullNP}
  \delta B(B,a) = \cP \left( \ln \frac1B - \cI(a) - \cJ(a) - 2 + \eta_0 +
    \psi(1+a) - \psi(1) + \frac{8^{a/2}-1}{a} \right).
\end{equation}
Finally, having evaluated the derivatives with respect to $a$ and
`undoing the expansion' with respect to powers of $1/Q$ we obtain
\begin{equation}
  \Sigma_N(B) = \Sigma_N^{(\mathrm{PT})}\left(B - \delta B(B,2R')\right)\,.
\end{equation}

In order to better understand our answer we shall consider two
important limits, $R'\to 0$ and $R'\to 2$.
\paragraph{Limit of $\boldsymbol{R'\to0}$.}

A useful cross check of the answer is to compare it with one's
expectations for $R'=0$. In this limit the perturbative broadening is
determined entirely by a single emission. Half the time
the emission will have been in the remnant hemisphere, implying $B =
p_t/Q$ and $\ln Q/p_t$ = $\ln 1/B$. The other half of the time the
emission will have been in the current hemisphere and $B=2p_t/Q$ and
so $\ln Q/p_t = \ln 2/B$. Taking the average one obtains
\begin{equation}
  \label{eq:deltaB0}
  \delta B(B,0) 
  = \cP\left(\ln
    \frac1B + \frac12 \ln 2 + \eta_0\right).
\end{equation}
Noting that $I(0) = \ln 2 - 2$ and $J(0) = 0$, we see that this agrees
with the full result, eq.~\eqref{eq:fullNP}.

\paragraph{Limit of $\boldsymbol{R'\to2}$.} A second limit of interest
is the point where $\Lambda$ diverges, $R'=2$. The techniques used
here for calculating the power correction are analogous to those of
section~\ref{sec:Derivation} for determining $\Lambda$. We know that
in the case of $\Lambda$ they break down pathologically for $R'\simeq2$.

In the case of $\delta B$ the method also breaks down, but in a less
pathological fashion. The reason is that $\delta B$ involves ratios of
divergent integrals, and so stays finite. In particular
\begin{equation}
  \cI(4) = 0,\qquad\quad \cJ(4) = 127\,,
\end{equation}
which leads us to the result that
\begin{equation}
  \delta B(B,4) = \cP\left(\ln \frac1B + \frac56 + \eta_0\right).
\end{equation}
In the case of $\Lambda$ we devoted considerable effort to obtaining a
correct answer in the region $R'=2$, even though this is somewhat to
the left of the peak.  Our motivation for doing this was two fold.
With more sophisticated models (\eg involving shape functions) for
non-perturbative effects, a perturbative understanding of that region
may still be of interest. Furthermore the method may be generalisable to
other observables for which the breakdown occurs much closer to the
peak (\eg for the difference between jet masses in $\ee$).

For the power correction however it is not clear that such an effort
is warranted: (a) the techniques that are required are probably more
complex than for $\LambdaBar$ (even without a specific treatment of
the $R'\simeq2$ region, the working for the power correction is somewhat
more complex than for the PT distribution); (b) in the region $R'=2$ the
simple approximation of a non-perturbative shift to the distribution
is in any case thought to be a poor approximation; (c) any techniques
developed would probably be useful only for the broadening.

Accordingly for phenomenology we advocate the use of
eq.~\eqref{eq:fullNP}. If one wishes to venture into the region around
$R'=2$ while bearing in mind that this is almost certainly not a safe
endeavour, one can ensure that the distribution remains well-behaved
by using the following extrapolation for $\delta B$ beyond $R'=2$:
\begin{equation}
  \delta B(B,a) = \cP\left(\ln \frac1B + \frac56 + \eta_0
    + 0.0795537 (a-4)
  \right), \qquad \quad a \ge 4\,,
\end{equation}
where the first derivative around $a=4$ has been determined
numerically.

Another region which deserves some discussion is that of large $B$.
Normally in $\ee$ the expression analogous to \eqref{eq:fullNP} is
used over the whole range of $B$. Of course for large $B$ the
expression is not valid, because it does not take into account
non-perturbative effects from a base configuration with 3 or more hard
partons. Nevertheless, phenomenologically this approach works rather
well as long as one does not go beyond the $3$-jet region, where the
distribution is in any case suppressed. In particular it is not
unreasonable to expect non-perturbative effects to shift the
distribution to the right even around the upper limit of the 3-jet
region, though perhaps not by exactly the same amount. This is because
the 3-jet upper limit is not the kinematical upper limit, and extra
soft gluon radiation is free to increase the value of the event shape.
(In the only case for which a calculation exists, the $C$-parameter at
the 3-jet limit, $C=3/4$, the power correction is found to be about
half that in the 2-jet region \cite{GPSZoltan})

In the case of $B_{zE}$ in DIS (as well as many other variables in DIS),
the situation is different --- the $2+1$-jet upper limit ($B=1/2$) is
also the kinematical limit for any number of particles and extra soft
radiation cannot increase the value of the event shape (though it can
reduce it). Accordingly it makes no sense to shift the distribution
by \eqref{eq:fullNP} around $B=1/2$. 

Of course we do not know what the right answer is. However one
solution, which does at least preserve the property that the
distribution should not extend beyond $B_{\mx} = 1/2$, is to replace 
\begin{equation}
  \label{eq:moddeltaB}
  \delta B  \to \delta \Btilde \equiv \left(1 -
    \left(\frac{B}{B_\mx}\right)^{p_{\mathrm{NP}}} \right) \delta B\,,
\end{equation}
where $p_{\mathrm{NP}}$ is an arbitrary positive power (which we would 
expect to take of order 1).  We refer to this
procedure as a \emph{modified} power correction, in analogy with the
modified matching of appendix~\ref{sec:ModMatch}, though we note that
the value of $p_\mathrm{NP}$ used for the power correction does not
have to be the same as $p$ used in the matching. As for the $p$ used
in matching it should be varied so as to gauge the systematic errors
associated with the arbitrariness of the procedure.

\subsection{Power correction to the mean}

We can use the above results to extract the power correction to
$\langle B_{zE}\rangle$. It can be written as
\begin{equation}
 \langle \delta B_{zE}\rangle(x,Q^2)  = 
 \int dB \frac{d\Sigma^{(\mathrm{PT})}(x,Q^2,B)}{dB} \delta B(B,2R'(B))\,.
\end{equation}
As can easily be seen, the integral converges for $\ln 1/B \sim \as^{-1/2}$.
Since our aim is only to control pieces down to a relative order of
$\sqrt{\as}$, it therefore is possible to make considerable simplifications
to both $\Sigma(x,Q^2,B)$ and to $\delta B$.  We use
\begin{equation}
  \Sigma^{(\mathrm{PT})}(x,Q^2,B) \to (1 + \asb  L {\cG}_{11}(x,Q^2)
  + \asb^2 L^3
  {G}_{23})\, e^{G_{12} \asb L^2}\,, \qquad L = \ln \frac1B\,,
\end{equation}
where ${\cG}_{11}$ is as defined in \eqref{eq:G11proj}, and
\begin{equation}
  \delta B(B,2R'(B)) \to \delta B(B,0)\,.
\end{equation}
We then exploit the following relations (dropping the
explicit $x,Q^2$ labels, for compactness)
\begin{align}
 \int dB \frac{d\Sigma^{(\mathrm{PT})}}{dB} &= 1\,, \\
 \int dB \frac{d\Sigma^{(\mathrm{PT})}}{dB} \ln \frac1B &= 
 \frac12\sqrt{\frac{\pi}{-G_{12}\asb}} - 
  \frac12\frac{{\cG}_{11}}{G_{12}} + 
 \frac12 \frac{G_{23}}{G_{12}^2} + \order{\sqrt{\as}}\,,
\end{align}
to obtain 
\begin{equation}
  \label{eq:meandeltaB}
  \langle \delta B_{zE}\rangle  = 
  \left(\frac14\sqrt{\frac{\pi}{\asb\CF}} +
  \frac34 - \frac{\bC_0 \otimes \bP \otimes \bq}{4\CF\, q} -
  \frac{\pi\beta_0}{3\CF} 
  + \eta_0 +\order{\sqrt{\as}}\right) \cP\,,
\end{equation}
where $\asb$ is to be evaluated at scale $Q^2$ (or $\mu^2$). We recall
that $\bq$ is the vector of quark and gluon distributions, with $q$
defined in \eqref{eq:qdef},
and that $\bP$ is the matrix of leading order splitting functions. An
alternative form for \eqref{eq:meandeltaB}, which may be more
practical to evaluate (but which introduces subleading corrections at
$\order{\as}$), is the following
\begin{equation}
  \label{eq:meandeltaBeasy}
  \langle \delta B_{zE}\rangle  = 
  \left(\frac14\sqrt{\frac{\pi}{\asb\CF}} +
    \frac34 - \frac{1}{4 \asb \CF} \frac{d \ln q}{d\ln Q^2}
  -\frac{\pi\beta_0}{3\CF} 
  + \eta_0 +\order{\sqrt{\as}}\right) \cP\,.
\end{equation}
If one determines the derivative of the quark distributions
numerically, then one should ensure that they are reasonably smooth in
$Q^2$, which as discussed in appendix~\ref{sec:PDF} is not always the
case.

We emphasise that the power correction acquires explicit $x$
dependence, through the dependence on the scaling violations of the
quark distributions. It is the first time that such a phenomenon is
seen for the mean value of DIS event shape.   It
would be of interest to make a comparison with the results of the ZEUS
collaboration, whose data seem to require non-negligible
$x$-dependence \cite{H1ZeusConf} in the power correction.

Finally we point out that the modification of the power correction
described at the end of the previous subsection, \eqref{eq:moddeltaB},
only affects the answer for the mean power correction,
\eqref{eq:meandeltaB}, at the level of terms $\order{\sqrt{\as}}$,
which are beyond our accuracy.

\section{Analysis of results}
\label{sec:Results}

Here we present numerical results based on the the calculations of the
previous sections. All figures have been generated with
$\as(M_Z)=0.118$ and where relevant
$\alpha_0(\mu_I=2\,\mathrm{GeV})=0.5$. Leading order parton evolution
has been used, consistent with the philosophy of keeping only leading
and next-to-leading logs in the resummation formula. The
renormalisation and factorisation scales have been kept equal to $Q$.

\subsection{Resummation versus fixed order}

\begin{figure}[tbp]
  \begin{center}
    \epsfig{file=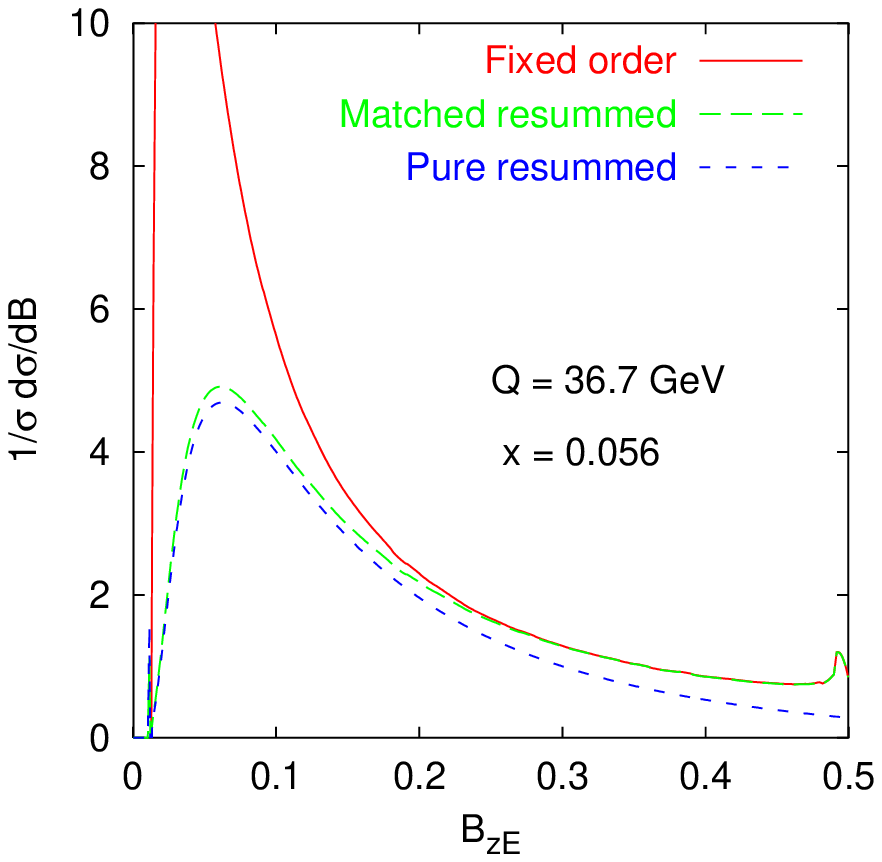,width=0.49\textwidth}\hfill
    \epsfig{file=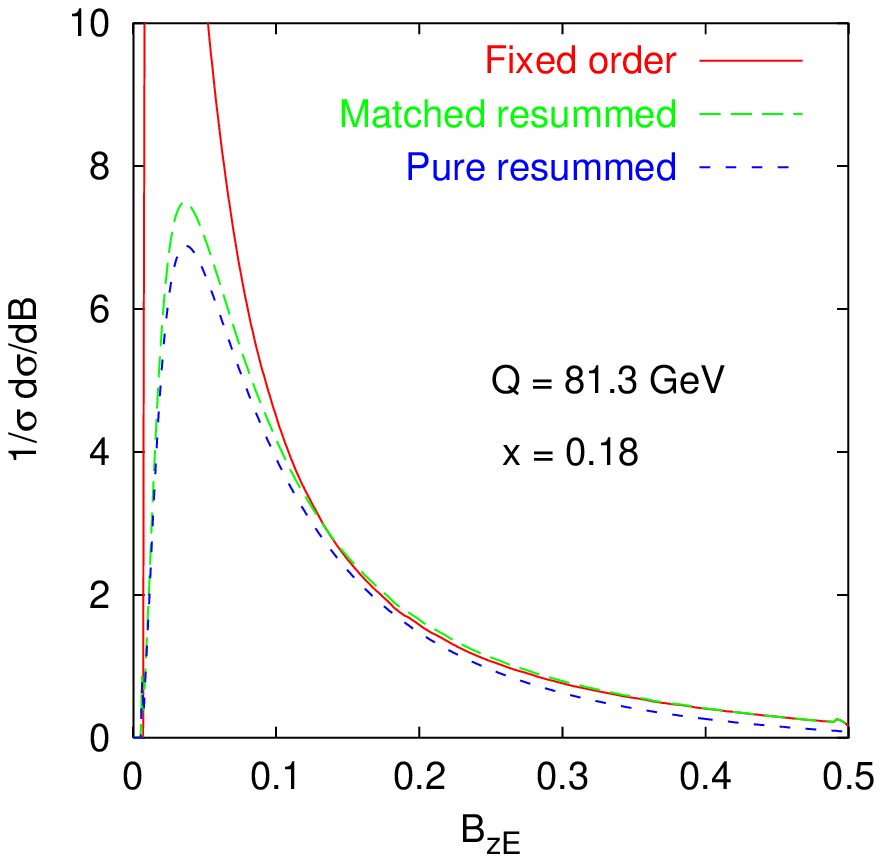,width=0.49\textwidth}
    \caption{A comparison of fixed order, resummed and
      matched-resummed distributions, at two $x,Q$ values for which
      measurements exist at HERA \cite{H1NewData}. The
      standard resummation has been used, and the matching is modified
      $M$-matching. No non-perturbative corrections have been
      included.}
    \label{fig:FOresum}
  \end{center}
\end{figure}

Figure~\ref{fig:FOresum} compares the resummed results (with and
without matching) to the fixed order ($\as^2$) prediction as
calculated with DISASTER++ \cite{Disaster}. The two plots correspond
to different $Q$ values. One sees that in the small-$B$ region the
resummation has a dramatic effect and that this effect is larger still
at lower $Q$ values. One also sees that the matched curves are
essentially identical to the fixed-order results at large $B$, while
at small $B$ the effect of matching amounts to a small modification of
the pure resummed results.

Close to the upper limit of the distribution one sees a small
secondary peak, most prominent at lower $x$ and $Q$ values. It seems
that this structure may actually be an artifact of the
interpolation of the fixed-order distribution, since there are arguments
that suggest that close to the maximum the distribution has an
integrable, $(\half-B)^{-1/2}$, divergence: when there is a single
particle in the current hemisphere, at an angle $\theta$ with respect
to the photon axis, the broadening is $B = \half \sin \theta$. For $\theta$
close to $\pi/2$ there is roughly a uniform distribution of $\theta$
values, then this translates into a $(\half-B)^{-1/2}$ behaviour for
the distribution of $B$. In practice this divergence will almost
certainly be smoothed out by soft gluon radiation, but that is beyond
the scope of this paper.

If we calculate the matched curves with DISENT \cite{CSDipole}
(the program is much faster, but is known to give wrong subleading
logarithms of $B$ \cite{DIStauz}) we find that the results are
modified by an amount of the order of a percent (for the lower $Q$
value). This small difference is a consequence of the matching that we
have used: in the small-$B$ region the matching terms are multiplied
by a Sudakov form factor, and therefore so are the discrepancies
in DISENT. If the matching term were not multiplied by a form
factor (as for example would be the case in $R$ matching) then the
discrepancy would be considerably larger, of the order of $6\%$ in the
peak region --- however given the difficulty of carrying out $R$
matching in DIS, this is unlikely to pose a problem.

Finally we note that at the higher $Q$ value shown there could be some
non-negligible contribution from $Z$-boson (rather than photon)
exchange. From the point of view of the resummation this makes no
difference, but for the non-logarithmically enhanced parts of the
fixed order calculation there could be an effect. Unfortunately of the
fixed-order programs that are able to calculate the broadening
distribution with reasonable accuracy, DISENT and DISASTER++, neither
implements $Z$-boson exchange.

\subsection{Resummation variants}

\begin{figure}[tbp]
  \begin{center}
    \epsfig{file=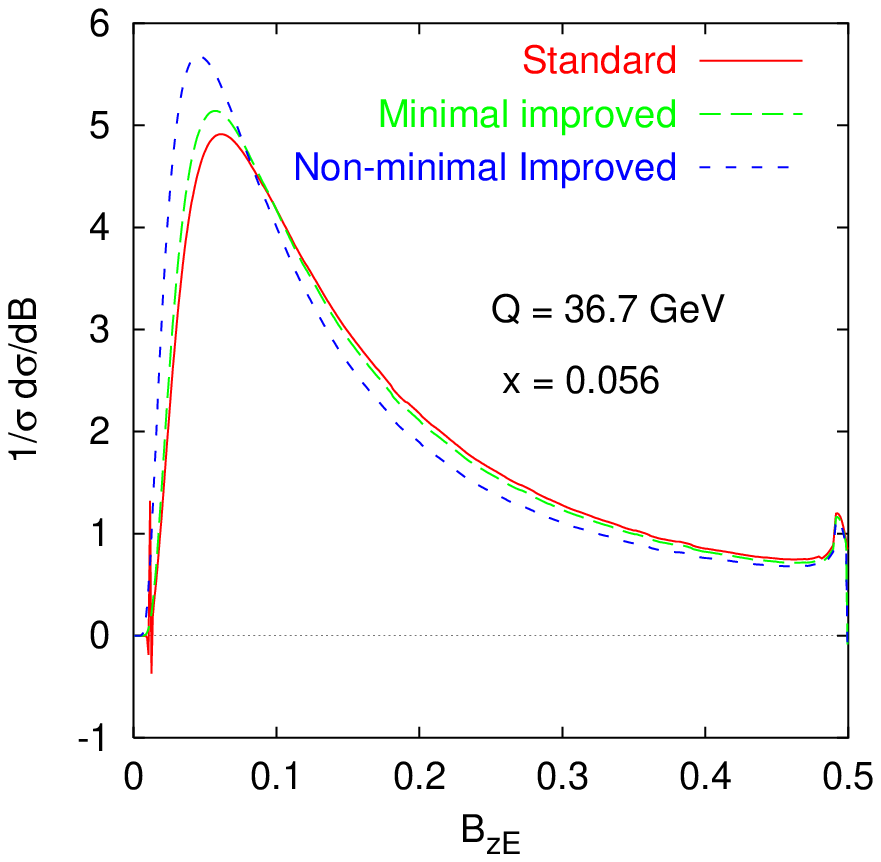,width=0.49\textwidth}\hfill
    \epsfig{file=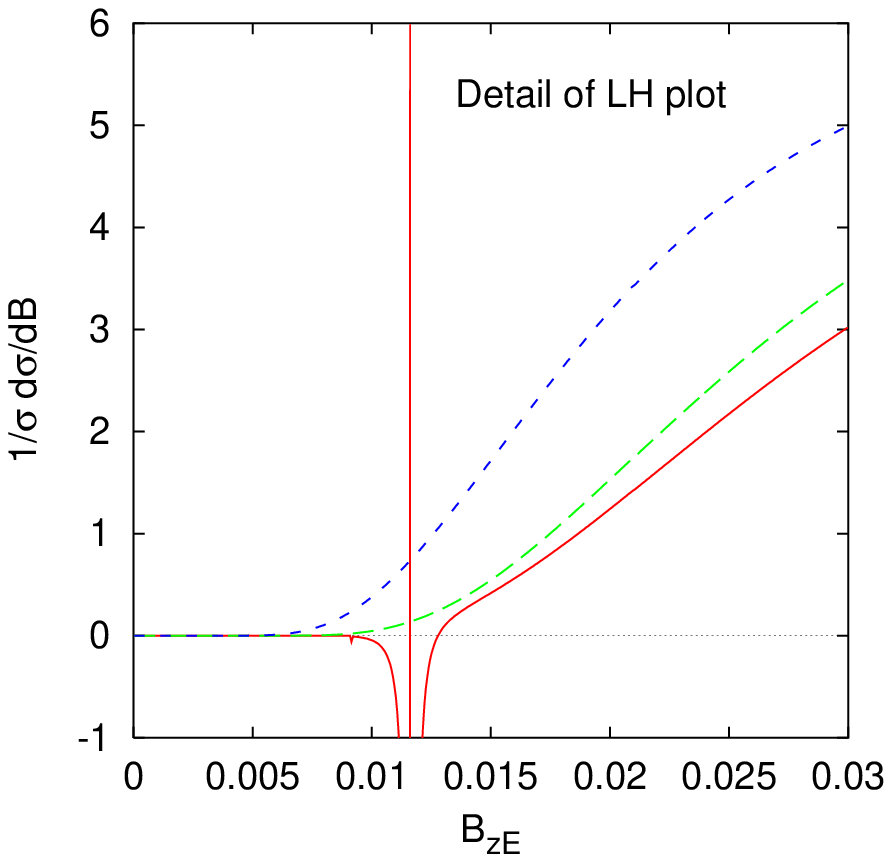,width=0.49\textwidth}
    \caption{A comparison of the standard, `improved' and `minimal
      improved' kinds of resummations. The right hand plot shows the
      same curves as the left-hand plot but with a magnified abscissa.
      All curves use modified $M$-matching and are shown without
      non-perturbative corrections. }
    \label{fig:3resum}
  \end{center}
\end{figure}

Figure~\ref{fig:3resum} shows the three different varieties of
resummation that have been developed in this paper. For the standard
resummation one sees some `noise' in the left-hand plot. In the
right-hand plot, which simply has a higher resolution on the $B$ axis,
one sees clearly the divergent structure associated with the
derivative of the pole in $\Lambda$ (which has been analytically
continued to give it meaning beyond $R'=2$). At lower $Q$ values the
problem is more prominent, but remains confined to a region which
though formally within perturbative reach, in practice is beyond the
applicability of the formulas.

We also show the two variants of the improved resummation. The minimal
improved curve is actually very close to the standard resummation,
while the non-minimal improved curve is somewhat different. This
difference is indicative of the size of uncontrolled subleading
effects. If one wishes to use an improved resummation we recommend the
`minimal' variant, mainly because of its similarity to the pure LL
plus NLL resummation. We note also that the non-minimal improved
resummation can show some small instabilities (not visible here) for
$R'\to\infty$, which 
arise in the derivative of the $((1 + \sqrt{1+y^2})/4)^{-R'}$ factor
in the integrand for $\LambdaBar_n$.

\subsection{Comparison with data}

\begin{figure}[tbp]
  \begin{center}
    \epsfig{file=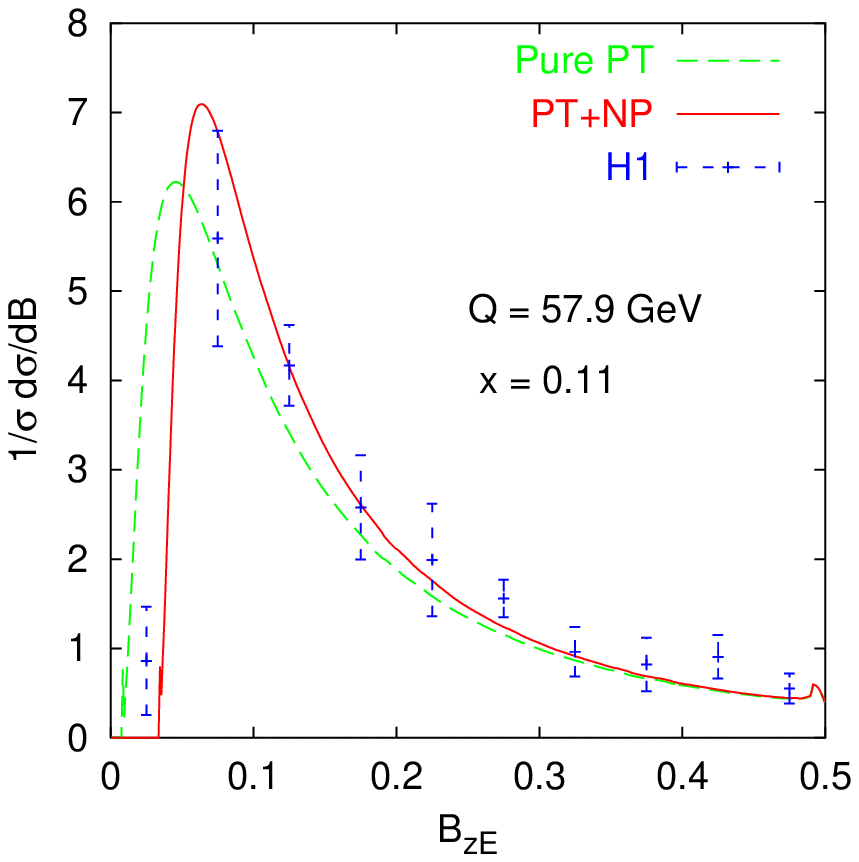,width=0.48\textwidth}\hfill
    \epsfig{file=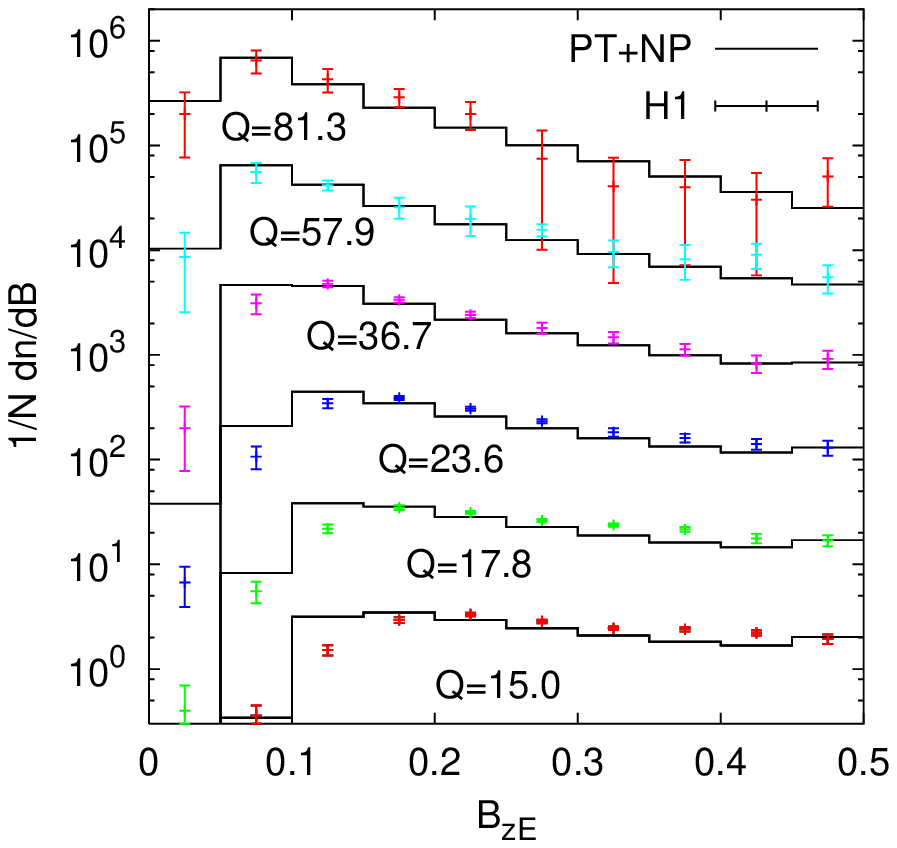,width=0.50\textwidth}
    \caption{Comparisons of the resummed results with H1 data
      \cite{H1NewData}. Experimental statistical and systematic, added
      in quadrature (and averaged if asymmetric). Note that in the
      right-hand plot the $x$ values are correlated with the $Q$
      values, going from $x=0.014$ at $Q=15$~GeV, to $x=0.18$ at
      $Q=81.3$~GeV, and that the distributions have been rescaled by a
      factor $10^i$ with $i$ running from $0$ to $5$ respectively.}
    \label{fig:data}
  \end{center}
\end{figure}

Figure~\ref{fig:data} shows the comparison of our results to some data
from H1 \cite{H1NewData}. The left hand plot shows a single $x,Q$
value, illustrating the pure perturbative result and the prediction
including the power correction (modified with
$p_\mathrm{NP}=2$). Because of the $\ln 1/B$ dependence, 
the effect of the power correction is not just to shift the prediction
but also to squeeze it, in a manner similar to what is seen for the
$\ee$ broadenings. Overall the description of the $x=0.11$ data is
reasonably good (we recall that we are using standard values of
$\as(M_Z)=0.118$ and $\alpha_0=0.5$). The apparently poor description
of the two lowest $B$ data points disappears in large part after
integration over the bin widths.

The right-hand plot of figure~\ref{fig:data} shows a comparison
between the data and the power-corrected resummations (now integrated
over bins) for a range of $x,Q$ values. There is a worsening of the
description at lower values of $Q$, but this is to some extent
expected, due to the increased relevance of subleading corrections
(both $1/Q^n$ \cite{KorchemskySterman} and higher orders in $\as$).

Elsewhere \cite{DSDIS} we shall present a more detailed analysis
together with that for a range of other variables, including a study
of renormalisation and factorisation scale dependence, of matching
scheme and rescaling dependence (\cfr appendix~\ref{sec:Rescaling}),
and of various other subleading effects.

\section{Conclusions}

For an accurate description of event shapes over the whole of the
available phase space it is necessary to carry out a resummation of
logarithmically enhanced terms and to supplement it with a
non-perturbative correction. This paper deals with the broadening
$B_{zE}$ and is one of a series addressing the resummation of a range
of event shapes in the Breit-frame current hemisphere in DIS
\cite{DIStauz,NonGlobal,DSDIS}.

In principle the broadening calculations involve a straightforward
extension of pre-existing resummation \cite{NewBroad,DIStauz} and
power correction \cite{BroadPower} techniques. In practice several
subtleties arise. The standard prescription of keeping just leading
and next-to-leading logarithms actually leads to a divergent answer.
This divergence is associated with a change in regime: at a certain
point as one goes to very small $B$, the coefficient of the double
logarithm is halved, because pure Sudakov suppression stops being the
favoured mechanism for producing the small $B$ values and is partially
replaced by a cancellation of recoil between multiple emissions.

In principle the divergence occurs inside the region which is supposed
to be under control. Accordingly we have developed techniques for
extending the resummation into the region with the divergence. It is
then necessary to modify our `accuracy criterion' --- rather than being
the correct treatment of LL and NLL terms, it becomes that for $\as
\ln 1/B \sim 1 $ the distribution should be under control up to (but
not including) corrections of relative order ${\as}$ (for normal
variables these two conditions are equivalent).

In practice the divergence lies in a region where relative corrections
of $\order{\as}$ are multiplied by such large (purely numerical)
coefficients that the distribution is in any case not well
constrained. Accordingly phenomenology is restricted to a region of
larger $B$, where it turns out to be sufficient to use the `standard'
resummation approach. We note however that the techniques developed
here may be applicable to other variables such as the difference between
jet masses in $\ee$ where a similar divergence is likely to be present
inside the phenomenologically relevant region.

We have also made other developments which are necessary for practical
phenomenology. In $\ee$ there exist two standard techniques for
matching the fixed order and resummed calculations. In DIS one of them
($R$-matching) is awkward to apply because of difficulties in
extracting the required fixed-order information from Monte Carlo
programs such as DISENT and DISASTER++. Another procedure, $\log R$
matching is applicable, but needs to be modified to address some
subtleties that arise in DIS but not in $\ee$. Furthermore we give
details on the matching procedure proposed in \cite{DIStauz}, naming
it $M$-matching, and also propose a new procedure, $M_2$ matching.

For the numerical implementation of our resummed formulae it
was useful to develop a software tool for the evolution of parton
distributions. A by product of this work was the discovery of bugs in
the NLO evolution embodied in the MRST (at small $x$) and CTEQ (in the
DIS scheme) parton distributions. In due course a stand-alone version
of the evolution program will be made public.

Another software development, which will be discussed in detail in
\cite{DSDIS}, is a tool which exploits factorisation to allow
fixed-order Monte Carlo events at a single value of $x,Q$ to be reused
for other $x,Q$ values. For DISASTER++ in particular this allows a
gain of an order of magnitude in speed and was essential for the
generation of the fixed order distributions used throughout this
paper.

With the availability of these tools we have therefore for the first
time been able to evaluate the resummed distribution of a DIS event
shape including NLO matching and compare the results to data. We find
reasonable agreement especially at higher $x$ and $Q$ values. More
detailed analysis, including the study of other variables, will be
presented in forthcoming work \cite{DSDIS}.

Finally we note that the programs to calculate the resummed broadening
distribution, including the matching and power correction, are
available from the following web page:
\texttt{http://cern.ch/gsalam/disresum/}.

\section*{Acknowledgments}

We are grateful to the Acharya-Boudjema Centre for Quantitative
Science for their generous hospitality while part of this work was
carried out. We would also like to thank Stefano Catani, Thomas Kluge,
Uli Martyn and Bryan Webber for useful conversations, and Dick
Roberts, James Stirling, Wu-Ki Tung and Andreas Vogt for numerous
helpful exchanges about parton distributions. We are also grateful to
the INFN, sezione di Milano and the Universities of Milano and Milano
Bicocca for the use of computing facilities.

\appendix

\section{Formulae for the radiators}
\label{sec:radiators}

We write the radiators eq.~\eqref{eq:RadiatorIntegrals} to NLL accuracy
as $R(1/B) = L R_1(\as L) + R_2(\as L)$ with
\begin{equation}
  R_1(\as L) = \frac{\CF}{2\pi\beta_0\lambda} \left(-2\lambda - \ln
    (1-2\lambda)\right) ,
\end{equation}
and
\begin{multline}
  R_2(\as L) = \frac{3\CF}{4\pi\beta_0} \ln(1-2\lambda)
     + \frac{\CF K}{4\pi^2 \beta_0^2}
            \frac{2\lambda +
              (1-2\lambda)\ln(1-2\lambda)}{1-2\lambda}\\ 
     + \frac{\CF \beta_1}{2\pi \beta_0^3}\left( -
       \frac{2\lambda + \ln(1-2\lambda)}{1-2\lambda} -
       \frac12\ln^2(1-2\lambda) 
     \right),
\end{multline}
where $\lambda = \as \beta_0 L$ and $L = \ln 1/B$. We have defined the
coefficients of the $\beta$-function to be
\begin{equation}
  \label{eq:betas}
  \beta_0 = \frac{11\CA - 2\nf}{12\pi},\qquad\quad 
  \beta_1 = \frac{17\CA^2 -5\CA\nf - 3\CF\nf}{24\pi^2}\,,
\end{equation}
and the constant relating the gluon Bremsstrahlung scheme \cite{CMW}
to the $\MSbar$ to be
\begin{equation}
K = \CA\left(\frac{67}{18} -
    \frac{\pi^2}{6}\right)
  - \frac{5}{9}\nf\,.
\end{equation}
For the radiator including the anomalous dimension, $R_\gamma$ we have
\begin{align}
  R_{\gamma1} &= R_1\,, \\
  R_{\gamma2} &= R_2 - \frac{\bgamma_N}{2\pi\beta_0} \ln(1-2\lambda)\,.
\end{align}
We also define
\begin{equation}
  \label{eq:Rp}
  R'(\as L) \equiv \frac{d}{dL} (L R_1(\as L)) = 
  \frac{\CF}{\pi \beta_0}  \frac{2\lambda}{1-2\lambda},
\end{equation}
and
\begin{equation}
  R'_2(\as, L) \equiv \frac{d}{dL} (R_2(\as L)) = 
  \frac{\as}{2\pi} \frac{-3\CF}{1-2\lambda}
  + \frac{\as}{2\pi} \frac{2K\beta_0 \CF  \lambda - 4\pi \beta_1 \CF \lambda
    \ln(1-2\lambda)}{\pi \beta_0^2 (1-2\lambda)^2} \,.
\end{equation}
The analogous results for the case with the anomalous dimensions are
\begin{align}
  R_{\gamma}' &= R' \\
  R_{\gamma2}' &= R_2' + \frac{\as}{2\pi}\frac{2\bgamma_N}{1-2\lambda}
\end{align}
Finally we shall need the leading parts of the second and third
derivatives:
\begin{align}
  R_\gamma'' \equiv R'' &= \frac{2\CF}{\pi} \frac{\as}{(1-2\lambda)^2}\,, \\
  R_\gamma''' \equiv R''' &= \frac{8\CF}{\pi}
  \frac{\as^2\beta_0}{(1-2\lambda)^3}\,. 
\end{align}

\section{Fixed order result}
\label{sec:FORes}

Here we calculate the first order coefficient function $\bC_1$, which
appears for example in \eqref{eq:xResum}. It is needed in a variety of
contexts --- for example in many (but not all) of the approaches to
matching with fixed-order calculations, and also for carrying out
comparisons down to $\asb^2 L^2$ accuracy with fixed-order
calculations.  Calculating $\bC_1$ proves more cumbersome in DIS than
in $e^{+}e^{-}$ since 
instead of a pure number 
one obtains a function of the 
variable $\xi = Q^2/(2p.q)$, where $p$ is the 
four-momentum of the incoming (as opposed to struck) parton. 
Additionally there are various
contributions to be computed, \ie transverse and longitudinal parts of
graphs with an incoming quark as well as boson gluon fusion.  It
proves convenient to first compute the $\mathcal{O}(\alpha_s)$ result
for events with broadening $B_{zE} > B$ because at leading order this
quantity, $\sigma_c^{(1)}$, doesn't get any virtual correction. Note
that this is complementary to the final quantity we want,
$\sigma_r^{(1)}$, which requires the selection of events with $B_{zE} <
B $.  The following relation is therefore required (which follows from
unitarity)
\begin{equation}
\label{subtr}
\sigma_r^{(1)} = \sigma_{\mathrm{tot}}^{(1)}  - \sigma_c^{(1)}\,,
\end{equation}
where $\sigma_{\mathrm{tot}}^{(1)}$ is the total
$\mathcal{O}(\alpha_s)$ cross-section for all events, except those
with an empty current hemisphere which are excluded throughout, since
it does not make much sense to define current jet observables in such
a situation.

The computation of $\sigma_c^{(1)}$ is relatively straightforward up to
terms of order $B$, which we do not require, and after applying
eq.~(\ref{subtr}) (at the level of the corresponding coefficient
functions ) we obtain the following pieces relevant to the computation
of $\sigma_r^{(1)}$

\begin{itemize}
\item $F_2$ quark contribution
\begin{multline}
\mathcal{F}_q(B,\xi) = -\delta(1-\xi) \left [4 \ln^2 B -3 \ln \frac{1}{B}
  -4\ln2 \ln B+2 \ln^2 2 +3 -3 \ln2-\frac{\pi^2}{3} \right ] \\ 
-2 \frac{1+\xi^2}{(1-\xi)_+} \ln \frac{1}{B} 
+\frac{1+\xi^2}{1-\xi} \ln{\xi} -(1+\xi^2) \left (
  \frac{\ln{(1-\xi)}}{1-\xi} \right )_+ +
\frac{6\xi^2-2\xi-1}{2(1-\xi)_+} \,.
\end{multline}
\item $F_2$ gluon contribution
\begin{equation}
\mathcal{F}_g(B,\xi) = -[\xi^2+(1-\xi^2)] \left [ 4 \ln \frac{1}{B}-2 +2 \ln
  \frac{1-\xi}{\xi} \right ]-8 \xi(1-\xi) \,.
\end{equation}
\end{itemize}
Note that the longitudinal contributions are absent since they cancel
in the two terms of eq.~(\ref{subtr}).  Then we have $\sigma_r^{(1)}$
is given by
\begin{equation}
\label{leadingord}
\sigma_r^{(1)}(x,B,Q^2) = \frac{1}{q(x)}\, \frac{\alpha_s}{2 \pi}
\int_x^1 \frac{d\xi}{\xi} \left [C_F
  \,\mathcal{F}_q(B,\xi)\,q(x/\xi)+T_f \,\mathcal{F}_g(B,\xi)
  \,g(x/\xi) \right ] \,,
\end{equation}   
with 
\begin{equation}
q(x) = \sum_{j=1}^{n_f} e_j^2 [q_j(x)+\bar{q}_j(x)] \, ,\quad T_f = T_R
\sum_{j=1}^{n_f} e_j^2 \, . 
\end{equation}
In the above $q_j(x)$ and $g(x)$ are quark and gluon distributions
(for a quark with flavour index $j$ and corresponding charge electric
charge $e_j$). The colour factors are as usual $C_F=4/3$, $T_R=1/2$
and $n_f$ denotes the number of active flavours.  From the above
leading order result the constant ($C_1$) and logarithmic ($G_{11}$
and $G_{12}$) pieces can be easily read-off.  In particular the
logarithms are in agreement with those obtained at this order from the
resummation method. 
The transpose of the matrix $\bC_1$ in section \ref{sec:MatchingMain} is  
\begin{equation}
 \bC_1^T(x) = \left(
    \begin{array}{c}
      e_u^2 C_{1,q}(x) \\ e_u^2 C_{1,q}(x) \\ \vdots \\ \sum_{q,\qbar} e_q^2
      C_{1,g}(x) 
    \end{array} 
     \right )
\end{equation}
with
\begin{subequations}
\begin{align}
  C_{1,q}(\xi) &= \CF \,{\mathcal F}_q (1,\xi)\,,\\
  C_{1,g}(\xi) &= \frac{T_R}{2}\, {\mathcal F}_g (1,\xi)\,.
\end{align}
\end{subequations}
Lastly we add that the above results are valid
for the DIS factorisation scheme.  To go the $\MSbar$ scheme one
should add the standard $\MSbar$ scheme $F_2$ coefficient functions to
the above results exactly as in eqs.~(A.9) and (A.10) in appendix A of
Ref.~\cite{DIStauz}.

\section{Rescaling the variable}
\label{sec:Rescaling}

Usually the resummation of a variable $V$ is defined in terms of $\ln
1/V$; however 
we can just as well define it in terms of $\ln 1/XV$ where $X$ is some
arbitrary number. If we do this then in the resummed formula we need
to make the following replacements (note that the meaning of overlined
symbols, such as $\Lovrln$, is not in any way related to the meaning of
barred symbols, $\Lbar$, introduced when discussing the improved
resummation):
\begin{align}
  L &\to \Lovrln = \ln \frac1{XV}\,,\\
  g_1(\as L) &\to g_1(\as \Lovrln)\,, \\
  g_2(\as L) &\to g_2(\as \Lovrln) +  (g_1(\as \Lovrln) + \as
  \Lovrln g_1'(\as \Lovrln)) \ln X\,,
\end{align}
where bold symbols for those functions that are
operators in $x$ and flavour space.

We also need to modify the constant term:
\begin{align}
  C_1 \to \Covrln_1 = C_1 + G_{12} \ln^2 X + G_{11} \ln X
\end{align}
In the fixed order expansion there are corresponding modifications which
need to be taken into account:
\begin{align}
  G_{11} &\to \Govrln_{11} = G_{11} + 2 G_{12} \ln X \\
  G_{22} &\to \Govrln_{22} = G_{22} + 3 G_{23} \ln X
\end{align}
which imply the following modifications at orders $\as$
\begin{equation}
  \sigma^{(1)}_r(L) \to \sigma^{(1)}_r(\Lovrln) + \Delta G_{11} L +
  \Delta C_1\,,
\end{equation}
and $\as^2$:
\begin{multline}
  \sigma^{(2)}_r(L) \to \sigma^{(2)}_r(\Lovrln) + 
  G_{12} \Delta G_{11} L^3 + \\ +
  \left(\Delta G_{22} + G_{11} \Delta
    G_{11} + \half (\Delta G_{11})^2 + G_{12} \Delta C_1 \right) L^2
  + \\
  + \left(G_{11} \Delta C_1 + \Delta G_{11} C_1 
          + \Delta G_{11} \Delta C_1   \right) L\,.
\end{multline}
We have defined $\Delta C_1 = \Covrln_1 - C$ and similarly
for other quantities. This expansion is valid in the case where
$C_1$ multiplies $q(V^n Q^2)$ rather than $q(Q^2)$ as was used in
\cite{DIStauz}. However the analogous expressions are
straightforward to determine for that case too.

The above formulae follow directly 
from the $N$ space resummed result but their 
re-interpretation, where required,  
as matrix projections 
in $x$ space is rather straightforward. 
We note that these formulae are currently not applicable to the
improved broadening resummations, which have further terms that need
to be taken into account.

\section{Accuracy checks}
\label{sec:accuracy}

In deriving the `improved' resummation in section~\ref{sec:Solution},
the stated aim was that the result should be correct to within
corrections of relative order $\as$. This requires a careful study of
contributions that have been neglected.

There are three main potential sources of inaccuracy that must be
considered:
\begin{itemize}
\item The choice of the expansion point.
\item The terms to be kept in the expansion of $R_\gamma$ (and whether
  they need to be kept in the exponent).
\item The choice of terms needed in evaluating the derivative of $\ln
  \sigma$, required for the inverse $\nu$-transform. 
\end{itemize}

\subsection{Choice of expansion point}

In the standard approach of section~\ref{sec:Derivation} the radiators
are both expanded around the point $\nu$. If one's only interest were to
obtain a convergent answer for $R'\gtrsim 2$ then this could be
achieved by keeping that expansion point and simply including the
second order expansion of $R$ in the exponent.

However for $R' > 2$ the saddle-point of the integral is far from
$bQ\sim \nu$. Indeed it is in a region where $\as \ln \nu/bQ \sim 1$,
and therefore in an expansion
\begin{equation}
  R(bQ) = R(\nu) + \sum_{n=1}^\infty R^{(n)} \left (\ln \frac{bQ}{\nu} \right )^n
\end{equation}
where $R^{(n)}(\nu)$ is of the form $\as^{n-1}f(\as L)$, it is
necessary to keep all terms. One way this could be achieved, is by not
expanding in the first place and integrating, say numerically, over
$b$ with the full function for $R(b)$. However this would lead to
problems because for sufficiently large values of $b$ one reaches a
singularity associated with the Landau pole in $R$.

If on the other hand we expand close to the saddle-point of the
integral the situation simplifies. The width of the integrand around
the saddle-point is roughly of order $1/\sqrt{R''} \sim \as^{-1/2}$,
allowing us to keep a fixed number of terms in our expansion.

For simplicity we choose to expand not around the actual saddle point,
but a point $b_0$ close to it, differing from it by a pure numerical factor.
Since in any case we will then have to keep terms in our expansion so
as to give an accurate representation of our function up to $\ln b/b_0
\sim \as^{-1/2}$, the difference of a pure factor between the
expansion and saddle points makes no difference.

\subsection{Choice of terms to be kept}

When examining the choice of terms to be kept, the discussion can be
kept simpler in the relevant region, $\as L \sim 1$, by noting that
any quantity which is formally $\as^n f(\as L)$ (with $f$ some
arbitrary function) is just of order $\as^n$. Accordingly we will
refer to $R''$ as being of order $\as$, $R'''$ as being of order
$\as^2$ and so on.

Keeping the first and second order terms in the expansion of $\cR_R$,
our basic integral is of the form
\begin{equation}
\lambda = \int
  \frac{dy}{y} \frac{y^2}{(1+y^2)^{3/2}} 
       \left(\frac{1+\sqrt{1+y^2}}{4}\right)^{-R'} 
       e^{-(\ln \frac{y}{2} - \ell)\Rpbar 
           -\frac12(\ln \frac{y}{2} - \ell)^2\Rppbar}\,.
\end{equation}
We see that for $\Rpbar \simeq 2$ it is the $\Rppbar$ term which
ensures the integral's convergence. It must therefore be kept in the
exponent. Examining the integral one sees that there are actually
three possible regimes. 
\begin{itemize}
\item   $2 - R' \gg \sqrt{\as}$. In this case the expansion point
  remains close to $y=1$ and the integral converges in a region of
  $\Delta \ln y$
  of order $1$.
\item $2 - R' \sim \sqrt{\as}$. In this case the saddle and expansion
  points remain close to $y=1$. The relevant integration region
  extends down to $\ln y \sim -\as^{-1/2}$, but convergence is rapid
  for $y > 1$.
\item $R'-2  \gg \sqrt{\as}$. Here $-\ell \gg \sqrt{\as}$ and
  accordingly the relevant part of the integral is entirely contained
  in the region $y \ll 1$ and all occurrences of $1 + y^2$ can simply
  be replaced by $2$.
\end{itemize}
The first region is simply that addressed in
section~\ref{sec:Derivation}, and one can neglect even the $R''$ term.

The second and third regions require more care. We need to understand
what happens when we multiply the integrand by a term $\as^m (\ln y -
\ell)^n$,
\begin{equation}
  \lambda_{mn} = \int
  \frac{dy}{y} \frac{y^2}{(1+y^2)^{3/2}} 
       \left(\frac{1+\sqrt{1+y^2}}{4}\right)^{-R'} 
       e^{-(\ln \frac{y}{2} - \ell)\Rpbar 
           -\frac12(\ln \frac{y}{2} - \ell)^2\Rppbar} \cdot \as^m
         \left(\ln \frac{y}{2} - \ell\right)^n\,. 
\end{equation}
Let us first consider the third region, where this simplifies to
\begin{equation}
 2^{2+R'} e^{2\ell} \int
  \frac{dy}{y} 
       e^{-(\ln \frac{y}{2} - \ell)(\Rpbar -2)
           -\frac12(\ln \frac{y}{2} - \ell)^2\Rppbar} \cdot \as^m
         \left(\ln \frac{y}{2} - \ell\right)^n\,. 
\end{equation}
If we are sufficiently into this region that we can neglect the
$(\Rpbar-2)$ term then we obtain
\begin{equation}
  \frac{\lambda_{mn}}{\lambda} \sim 
  \left\{
  \begin{array}{ll}
    \as^{m-n/2} & \mbox{$n$ even}\\
    0           & \mbox{$n$ odd}
  \end{array}\right.
\end{equation}
The largest such contributions will come from terms such as
$(R_{\gamma 2}')^2$, ${R'''}^2$, $R^{(4)}$, and will all be of
relative order $\as$, and so negligible.

In the situation where $2 - R' \sim \sqrt{\as}$ the situation is more
complex because the odd-$n$ terms do not give zero. The reason is that
the integral extends only to one side of the saddle-point so one loses
the cancellation between $\ln y/2 - \ell > 0$ and $\ln y/2 - \ell <
0$. Accordingly, in this region
\begin{equation}
  \frac{\lambda_{mn}}{\lambda} \sim 
  \as^{m-n/2}, \qquad n \ge 0\,.
\end{equation}
Accordingly we must keep all terms $\as (\ln y/2 - \ell)$ and 
$\as^2 (\ln y/2 - \ell)^3$, since they contribute at the relative
$\order{\sqrt{\as}}$
level. This is the motivation behind the set of terms kept in
eq.~\eqref{eq:expansion}.

\subsection{Terms to be kept in the derivative of
  $\boldsymbol{\LambdaBar}$} 

In evaluating the inverse Mellin transform with respect to $\nu$, it
is necessary to calculate the factor
\begin{equation}
  \label{eq:GammaDeriv}
  \frac{1}{\Gamma\left(1 + \nu\frac{d}{d\nu} \ln \sigma\right) }\,.
\end{equation}
In the second and third regions discussed above $\LambdaBar$ varies
rapidly, so its derivative can contribute significantly to this factor
and should not be neglected.

The derivative of the full $\ln \LambdaBar$ is technically quite
complicated to evaluate because of the presence of the
anomalous-dimension matrices. However these terms, which are of order
$\as (\ln y/2 - \ell)$, have two important features: they give a
contribution of relative order $\sqrt{\as}$, and that contribution is
relevant (and varies significantly) only in the region of $\Delta L
\sim \as^{-1/2}$. Consequently in $\ln \LambdaBar$ when taking the
derivative with respect to $L$, these extra pieces become of order
$\as$ rather than $\sqrt{\as}$, and so can be neglected. The same is
true of all terms which contribute a relative amount $\sqrt{\as}$ in
this limited region of $L$.

The fact that we can throw away terms contributing a relative amount
$\sqrt{\as}$ in that region also allows us to modify the large-$y$
structure of the integrand, and therefore we can use $\LambdaBar_m$ rather
than $\LambdaBar_n$. That the large-$y$ region contributes an amount
of relative order $\sqrt{\as}$ follows from the fact that the whole
integral is of order $\as^{-1/2}$, while the large-$y$ region is
unenhanced and contributes an amount of order $1$.

\section{Modified matching}
\label{sec:ModMatch}
As was discussed in section \ref{sec:MatchingMain}, it is important that 
the matching respect certain properties concerning the behaviour at 
the maximum of the distribution $V = V_\mx$. 
We recall that these were: 
first the integrated cross section should go to whatever the
correct upper limit happens to be, without leftover terms of order
$\as^3$ or higher. Secondly if the fixed order distribution
goes to zero smoothly at the upper limit, so should the
matched-resummed one. In $\ee$ this was always the case, whereas for
many DIS variables the distribution is non-zero at the upper limit
(and the upper limit is the same for all orders). Here we discuss a 
modified matching procedure 
that makes sure that the final answer has the required 
behaviour in the above respects.

The first element of the modification is to replace\footnote{Certain
  resummations are defined, by default, not in terms of $\ln 1/V$ but
  rather in terms of $\ln V_0/V$. One such example is the
  $C$-parameter with $V_0=6$, both in $\ee$ \cite{Cpar} and in DIS
  \cite{DSDIS}. We write our formulae so that they are valid in these
  cases too.}
\begin{equation}
  L = \ln \frac{V_0}{V} \to \Ltilde = \frac{1}{p} \ln \left[
    \left(\frac{V_0}{V}\right)^p - 
    \left(\frac{V_0}{V_\mx}\right)^p + 1\right]\,,
\end{equation}
where we have generalised through the inclusion of the power $p$ the
modification originally proposed in \cite{CTTW} (which used $p=1$).
Typically one might expect to consider $ 1 \le p \lesssim 3$, where the
upper limit is fairly arbitrary and the lower one comes from an
assumption that the cross section contains no terms of the form $V^p
\ln V$ with $p < 1$.  In cases where we use a rescaled variable then
we have
\begin{equation}
  \Lovrln = \ln \frac{V_0}{X V} \to 
  \Lovrlntilde = \frac{1}{p}
  \ln \left[ 
    \left(\frac{V_0}{X V}\right)^p - 
    \left(\frac{V_0}{X V_\mx}\right)^p + 1\right]\,.
\end{equation}
We shall also need a factor with the property that it goes to $1$ rapidly
for $V\to0$ and to zero for $V = V_\mx$. We adopt the following form
for it:
\begin{equation}
  Z(V) = 1 - \left(\frac{V}{V_\mx}\right)^p.
\end{equation}

For $M$ and $M_2$ matching, the replacement of $L$ with $\Ltilde$ is
usually sufficient to ensure that cross section goes exactly to the
$\order{\as^2}$ upper limit. This is because at $V = V_\mx$, $ \Ltilde=0$,
and accordingly $\sigma_r$ contains no terms higher than
$\order{\as}$, and the matching adjusts both the order $\as$ and
$\as^2$ terms to be exact, without introducing any additional terms.
There is an exception to this rule for the improved broadening
formula, which contains $\order{\as^3}$ terms at $V_\mx$ even after
the replacement $L \to \Ltilde$ --- these then need to be subtracted.

If $\sigma^{(2)}_e(V_\mx)$ is non-zero (as it usually is), then the
matched \emph{distribution} does not go to zero even if the fixed
order one does, because of the $\asb^3 \sigma^{(2)}_e(V_\mx) \cG_{11}$
contribution. Technically the most straightforward solution to the
above two problems is to use the following formula
\begin{multline}
  \sigma(V) = \sigma_r + \left[\asb \left(\sigma_e^{(1)} -
    \sigma_r^{(1)}\right) + \asb^2 \left(\sigma_e^{(2)} -
    \sigma_r^{(2)}\right) - 
  \asb^2 Z(V) \left(\sigma_e^{(1)}-
    \sigma_r^{(1)}\right)(L^2 G_{12} + L \cG_{11})
\right. \\ \left.
   + 1 + \asb \sigma_r^{(1)}(V_\mx) + \asb^2 \sigma_r^{(2)}(V_\mx) 
   - \sigma_r(V_\mx)
\right] \Sigma^{Z(V)}\,,
\end{multline}
though strictly speaking only the $G_{11}$ part of $\Sigma$ need be
raised to the power $Z(V)$.

Correspondingly for $M_2$ matching we have
\begin{multline}
  \sigma(V) = \sigma_r + \asb \left(\sigma_e^{(1)} -
    \sigma_r^{(1)}\right) + \left[ \asb^2 \left(\sigma_e^{(2)} -
    \sigma_r^{(2)}\right)
  \right. \\ \left.
    + 1 + \asb \sigma_r^{(1)}(V_\mx) + \asb^2 \sigma_r^{(2)}(V_\mx) 
   - \sigma_r(V_\mx)    
    \right]
\Sigma^{Z(V)}\,.
\end{multline}

For $\ln R$ matching the situation is different in that, following the
replacement $L \to \Ltilde$, the matched distribution goes to zero if
the fixed-order one does, but that now we need to fix up the value of
the cross section at the upper limit. In this respect the procedure
differs from the $\ee$ case, where replacing $L \to \Ltilde$ led to
all required properties automatically being satisfied. There are two
reasons for the difference: the fact that we keep the full
$\sigma_{rq}$ in front of the exponential (in $\ee$ the constant part
is left out) and the fact that the fixed order cross section does not
go to exactly $1$ at the upper limit, but to $1 + \order{\as}$. To
ensure that we get exactly the same answer as the $\order{\as^2}$
answer we need to insert an extra factor $F$ in front of the
exponential:
\begin{equation}
  \sigma(V) = \sigma_{rg} + 
  \sigma_{rq} \,F\, e^{\asb\left(\sigma^{(1)}_e - \sigma^{(1)}_r\right)
     + \asb^2\left(\sigma^{(2)}_e - \sigma^{(2)}_r 
       -\frac12\left(\sigma^{(1)}_e - \sigma^{(1)}_r\right)
       \left(\sigma^{(1)}_e + \sigma^{(1)}_r - 2\sigma^{(1)}_{rg}\right)
      \right)}\,,
\end{equation}
with
\begin{multline}
  F = 
  e^{ 
    -\asb\left( \sigma^{(1)}_{e} - \sigma^{(1)}_{r}\right)
    - \asb^2\left(\sigma^{(2)}_{e} - \sigma^{(2)}_{rq} 
      - \frac12\left(\sigma^{(1)}_{e} -
        \sigma^{(1)}_{r}\right)
      \left(\sigma^{(1)}_{e} + \sigma^{(1)}_{r}
         - 2\sigma^{(1)}_{rg}
      \right)
    \right)
    }
  \\
 \left.
   \frac{
    1 + \asb(\sigma^{(1)}_{e} - \sigma^{(1)}_{rg}) +
    \asb^2\sigma^{(2)}_{e}} 
  {1 + \asb \sigma^{(1)}_{rq} + \asb^2 \sigma^{(2)}_{rq}}
  \right|_{V=V_\mx}
  \,.
\end{multline}
where have exploited the fact that $\sigma^{(2)}_{rg}(V_\mx)$ is zero.
Since $F = 1 + \order{\as^3}$ the inclusion of the factor $F$
makes no difference (at our accuracy) to the small-$V$ behaviour of the
answer.

To see that the matched distribution goes to zero if the fixed-order
one does, we restrict ourselves to the case of variables without
any variable dependent scale 
in the structure function, because it turns out that
none of the variables involving $V$ in the scale of the distribution
have fixed-order distributions which go to zero at the upper limit. So
with this proviso one recalls
\begin{equation}
  \sigma_{rq} = (1 + \asb {\mathcal C}_1) e^{\Ltilde g_1(\as \Ltilde)
    + g_2(\as     \Ltilde)}\,\qquad\quad 
{\mathcal C}_1 = \frac{\bC_1 \otimes \bq}{q}\,.
\end{equation}
In the exponent, after matching, the $\as$ and $\as^2$ will have
vanishing derivatives by construction (\ie from the inclusion of the
fixed order pieces). At order $\as^3$ we get contribution only from
$g_1$ and $g_2$, but they go as $\Ltilde^4$ and $\Ltilde^3$
respectively and accordingly their derivatives vanish. This ensures
that the quark resummation part gives a vanishing distribution.

We also need to show that the gluon resummation piece gives a
vanishing distribution --- this follows since, because we have
included no $g_2$ term in the gluon resummation exponent, the lowest
power of $\Ltilde$ that is present is $\Ltilde^2$, automatically
giving zero derivative.

\section{Parton distributions}
\label{sec:PDF}

A complication in the practical computation of the
broadening (and also $\tau_{zE}$ and $\tau_{zQ}$) distribution in DIS
arises from the 
fact that the main result \eqref{eq:StandardResult} involves operators
in $x$ and flavour space, a consequence of the presence of the
anomalous dimension matrix, $\bgamma_N$,  in $R_\gamma$.

The action of $\bgamma_N$ is of course just to change the scale of the
parton distributions to $(VQ)^2$, \ie we can rewrite
\eqref{eq:StandardResult} as
\begin{equation}
  \label{eq:StandardResultExp}
  \Sigma_N(B) =   \bC_{0,N} \frac{\Lambda(2R')}{\Gamma(1 + 2 R')}
  e^{-2R(1/B) - R' (\ln 2 + 2\gamma_E) }
  \boldsymbol{q}_N(V^2Q^2)
\,.
\end{equation}
Parton distributions are available in tabulated form as a function of
scale from groups such as MRST \cite{MRST}, CTEQ \cite{CTEQ} or GRV
\cite{GRV}, so we could just choose to use these tabulated
distributions at scale $V^2Q^2$. This is what is usually done in the
context for example of Drell-Yan $p_t$ resummations (for recent
examples, see \cite{DYPairRecent} and references therein).

Instead however we have chosen to take a seemingly more complicated
route and develop our own software for the evolution of parton
distributions. There are four main reasons for this.

\paragraph{Matching.} 
When we carry out matching to fixed-order calculations we need to know
quantities such as
\begin{equation}
  \label{eq:MatchingQuants}
  \bP^{(1)} \otimes q\,,\qquad  \bP^{(2)} \otimes q\,,\qquad
  \bP^{(1)} \otimes\bP^{(1)} \otimes q\,,
\end{equation}
where $\bP^{(1)}$ and $\bP^{(2)}$ are the matrices respectively of
leading and next-to-leading order splitting functions. By taking
numerical derivatives it would be possible to obtain certain
combinations of the above quantities, for example
\begin{equation}
  \left(\asb \bP^{(1)} +  \asb^2 \bP^{(2)} \right) \otimes q
\end{equation}
from the first derivative. However this combination would be obtained
only for a particular value of $\as$ (that used in the original
evolution). The double convolution term can be obtained from the
second derivative of the structure functions, but only in a form
`polluted' by additional $\order{\as^3}$ terms. Furthermore some
current tabulated (global fit) parton distribution sets use quite poor
interpolation which means that numerically determined derivatives
(especially higher derivatives) are nonsensical.

So we in any case need to write software to calculate the quantities
in \eqref{eq:MatchingQuants}. Once this is done, writing a full PDF
evolution program involves relatively little extra work.

\paragraph{Flexibility.} 
If we follow the philosophy of $\ee$ event-shape resummations, \ie we
include LL and NLL terms but no higher-order contributions (other than
as introduced in the matching), then we must take
eq.~\eqref{eq:StandardResult} literally, using only the leading-order
splitting functions. If on the other hand we use
\eqref{eq:StandardResultExp} with tabulated global-fit parton
distributions then 
the evolution embodied in $q(V^2 Q^2)$ will automatically include NLO
splitting functions (either way we have to use NLO parton densities
since we match to $\order{\as^2}$ fixed order calculations).

Using our own evolution code gives us a certain flexibility, and we
are free to take a NLO parton distribution at scale $Q$, and then
apply to it the operator $\exp(-R_\gamma(B))$, \ie carry out leading
order evolution to scale $V Q$.

\paragraph{Fitting $\boldsymbol{\as}$.}
When fitting for $\as$, for each new value of $\as$ that one wishes to
examine, the formally correct procedure is to reevaluate all
quantities using parton distributions fitted with that value of
$\as$.

With the tools that are currently available for calculating
fixed-order distributions, this means
rerunning DISENT or DISASTER++, a process which can use several tens of
days of computing time on a modern workstation. As a result it is
common to fit for $\as$ using a single PDF set and then to check that
the results do not change significantly with a set corresponding to a
different value of $\as$.

\begin{figure}[ht]
  \begin{center}
    \epsfig{file=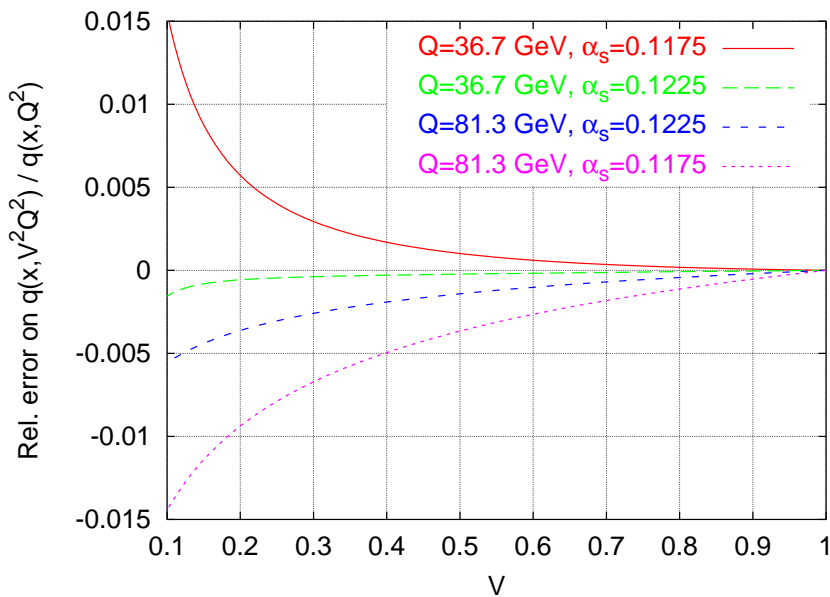,width=0.7\textwidth}
    \caption{The relative error on the ratio \eqref{eq:ratios}
      introduced by (improper) use of the central MRST PDF set to calculate
      quantities at $\as(M_Z)=0.1225$ (where one should formally use
      the high-$\as$ MRST distribution). Shown for
      $x=0.056,\,Q=36.7$~GeV and $x=0.18,\,Q=81.3$~GeV. The value of
      $\as$ is that used to evolve the central MRST parton
      distribution down from scale $Q^2$ to scale $V^2 Q^2$.}
    \label{fig:EvlnVarAs}
  \end{center}
\end{figure}

Within this approach, if we insert tabulated global-fit distributions into
\eqref{eq:StandardResultExp}, then while most of the calculation will
be done with the value of $\as$ that is explicitly inserted into the
formulas, the single logs associated with the anomalous dimension will
be evaluated with a value of $\as$ corresponding to the PDF set. We
can study the extent to which this is a problem by examining how the
ratio
\begin{equation}
  \label{eq:ratios}
  \frac{q(x,V^2 Q^2)}{q(x,Q^2)}
\end{equation}
depends on different approaches used for its evaluation (we recall our 
earlier definition for $q$, eq.~\eqref{eq:qdef}). Suppose we are
calculating 
the broadening distribution with $\as(M_Z)=0.1225$.  The correct
procedure is to calculate the ratio \eqref{eq:ratios} using the MRST99
hi-$\as$ ($\as(M_Z)=0.1225$) set evolved down from $Q$ to $V Q$ with
this appropriate value of $\as(M_Z)$.  We want to examine the error
that is made if one calculates this ratio in two different ways (as
would be done in a fit): (a) with the central-$\as$ MRST99 (originally
fitted with $\as(M_Z)=0.1175$) set evaluated at scale $Q^2$ and
evolved down to $V^2 Q^2$ with $\as(M_Z) = 0.1175$ --- this is roughly
equivalent to using the `central' tabulated global-fit distribution in
both the 
numerator and denominator of \eqref{eq:StandardResultExp}; (b) with
the central-$\as$ MRST99 set evaluated at scale $Q^2$ and evolved down
to $V^2 Q^2$ with $\as(M_Z) = 0.1225$ --- this is the philosophy of
treating the anomalous dimension terms on the same footing as all the
other single logs. The error that arises in these two different
approaches is shown in figure~\ref{fig:EvlnVarAs}, where one sees that
the $\as$ used in the evolution is considerably more important than
the $\as$ used in the fit for the parton distributions.

It should however be noted that for small-$x$ and small-$Q$ we can
have the opposite situation because of the strong correlation between
the value of $\as$ and the fitted gluon distribution.
Nevertheless this analysis indicates that at the very least we want
the option of doing our own evolution of the parton distribution.

\paragraph{Technical limitations of tabulated global-fit distributions.} 

One final motivation for using our evolution in calculating is $q(V^2Q^2)$
is smoothness. Generally global-fit parton distributions are provided in
tabulated form together with an interpolating program.

\begin{figure}[ht]
  \begin{center}
    \epsfig{file=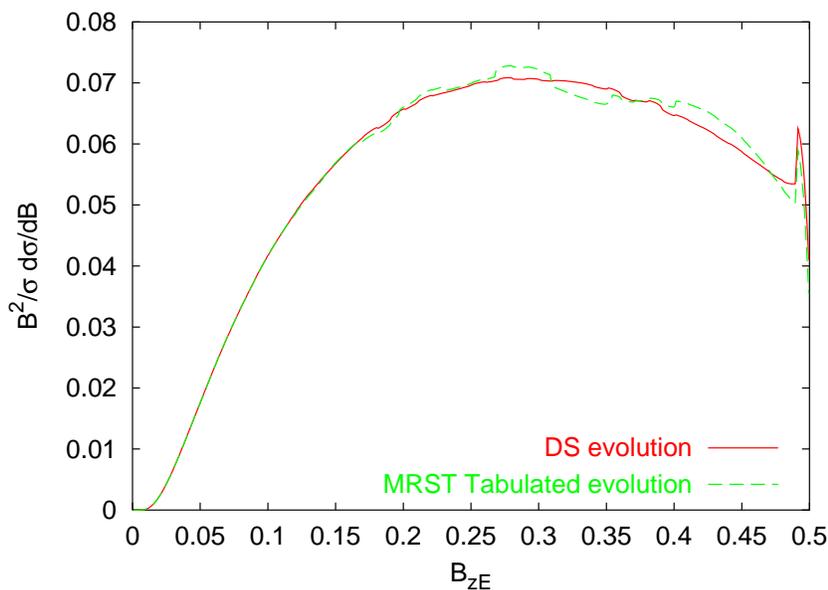,width=0.7\textwidth}
    \caption{The broadening distribution (without power correction)
      calculated using our own (DS) evolution and with the MRST99 tabulated
      evolution. Shown for $x=0.18,\, Q=81.3$~GeV. For other $x,Q$
      points of interest at HERA, the non-smoothness of the MRST
      distributions has a smaller effect. A leading factor of $B^2$
      has been included so as to highlight the large-$B$ region, where 
      the problem is more severe.}
    \label{fig:PRMvNLO}
  \end{center}
\end{figure}

In the case of the CTEQ and GRV distributions the interpolation is of
reasonable quality. However for the current publicly available MRST
distributions (dating from 1999, \cite{MRST}) only linear
interpolation is used, leading to non-smoothness in $Q$. When
calculating a resummed distribution (as opposed to the integrated
cross section) one takes the derivative of
\eqref{eq:StandardResultExp} and this non-smoothness gets promoted to
discontinuities. This is illustrated in figure~\ref{fig:PRMvNLO} which
shows the broadening distribution for $Q=81.3$~GeV, determined in two
ways: in one case we have used our own evolution between scales $Q$
and $VQ$; in the other case we have used the MRST99 tabulated
distributions to obtain $q(V^2Q^2)$. The clear discrepancies between
the two curves (at the level of a few percent) are a consequence
of the non-smoothness of the tabulated distribution.

We have also tested a preliminary version of the MRST 2001
distributions \cite{MRST2001}, which use improved interpolation code,
and there we find that the problem is eliminated.

\subsection{Convolution and evolution algorithms.}

We have seen above that there are several motivations for evolving the
parton distributions independently from the original global-fit tabulations.
Accordingly we have developed our own evolution and convolution code.
Various requirements arise from the need to use it for event shape
resummations (though we envisage that it may well have wider
applications):
\begin{itemize}
\item Flexibility: it should be straightforward to implement new
  kernels (\ie without any special analytical work), since each event
  shape involves a new constant piece $C_1$. This also makes it
  straightforward to extend it to NNLL evolution.
\item Reasonable tradeoff between speed and accuracy: both setup and
  evolution should be relatively quick.
\item Robustness: we will be evolving tabulated distributions, which
  may have imperfections due to interpolation, and which are not
  usually available in analytic form. Also, given that one may at some
  stage wish to use the code in a very different context (\eg
  evolution of fragmentation functions to scales of interest for
  high-energy cosmic rays \cite{CosmicRays}) we prefer to avoid any
  reliance on certain properties of `nice behaviour' of the structure
  functions and their derivatives as in \cite{Kosower} may reduce the
  robustness of the algorithm (though it can enable a considerable
  increase in speed).
\end{itemize}

The code has been written with a modular, semi object-oriented
approach, within the limits of the programming language used,
Fortran~90. It uses an optimised $x$-space algorithm (as opposed to 
$N$-space, where $N$ is the Mellin transform variable conjugate to
$x$). Parton distributions are represented on a grid with $n$ points
uniformly spaced in $\ln 1/x$:
\begin{equation}
  q(x) \Rightarrow q_i \equiv q(x_i)
\end{equation}
where $x_i = \exp(-i\delta)$ with $\delta$ the grid spacing. The
parton distribution at arbitrary $x$ is then defined to be equal to a
linear combination of the parton distribution on neighbouring grid
points
\begin{equation}
  \label{eq:PDFInterp}
  q(x) = \sum_{j=i(x)}^{i(x)+p} c_j(x) q_j  \,.
\end{equation}
where $i(x)$ is a grid point close to $x$. The coefficients
$c_j(x)$ are chosen so that $q(x)$ corresponds to the
$p^{\mathrm{th}}$ order interpolation of the points $i(x)$ to
$i(x)+p$. If one chooses
\begin{equation}
  \label{eq:InterpPosn}
  i(xe^{-\delta}) = i(x) + 1\,,
\end{equation}
then one has the
property
\begin{equation}
  \label{eq:NiceProp}
  c_j(x) = c_{j+1}\left(xe^{-\delta}\right)\,.
\end{equation}

With this representation of the parton distribution, convolutions then
become sums:
\begin{equation}
  \label{eq:PDFConv}
  [P \otimes q](x) = \int_x^1 \frac{dz}{z} P(x/z) q(z)
  \Rightarrow [P \otimes q]_i = \sum_{j=0}^i P_{ij} q_j\,,
\end{equation}
where
\begin{equation}
  P_{ij} = \int^{1}_{x_i} \frac{dz}{z} P(x_i/z) c_j(z)\,.
\end{equation}
Approaches of this kind have been adopted by many people, for example
\cite{Botje,Santorelli:1999yt,Zomer,Ratcliffe}, and indeed are a
standard numerical method \cite{NumRec}. In the algorithms of
\cite{Botje,Santorelli:1999yt} the grid is non-uniform in $\ln x$ and
there are roughly $n^2/2$ elements of the $P_{ij}$. Their
evaluation leads to considerable overheads (analytical or numerical
depending on the approach), while the requirement
that they be held in memory during program execution (rather than on
disk) places an upper limit on the value of $n$ that can be used.

As has been exploited by \cite{Zomer,Ratcliffe}, on a uniform grid,
away from the edges of the integral, the property \eqref{eq:NiceProp}
allows a further simplification:
\begin{equation}
  \label{eq:PijProps}
  P_{ij} = P_{(i+1)(j+1)}\,,
\end{equation}
\ie $P_{ij}$ is only a function of $i-j$. Thus one has only $n$
elements to evaluate and store, making it feasible to go to large values of
$n$. Furthermore the operators for multiple
convolutions are straightforward to evaluate, for example
\begin{equation}
  P^2_{ik} = \sum_j P_{ij} P_{jk}
\end{equation}
and $P^2_{ik}$ can be completely determined with $\order{n^2}$
operations. In contrast if \eqref{eq:PijProps} does not hold then one
needs $\order{n^3}$ operations.

In \cite{Zomer} this method has been implemented only for $p=1$, while
\cite{Ratcliffe} has applied it to the general $p$ case. For the
general-$p$ case some subtleties arise with grid edges, because
eqs.~\eqref{eq:PDFInterp} and \eqref{eq:InterpPosn} taken together
imply a sum over points outside one's grid. The approach in
\cite{Ratcliffe} is to sacrifice the formal accuracy of the approach
at large $x$, so that for the integration region between grid points
$j-1$ and $j$, the order is $\min(p,j)^\mathrm{th}$ order.

Here we observe that by choosing the $i(x)$ appropriately (and
differently according to the value of $x$ used in \eqref{eq:PDFConv})
one can maintain the full $p^\mathrm{th}$ order accuracy of the
method. The price that we pay is that $P_{ij}$ now requires the
computation of $\order{pn}$ entries rather than $n$, the extra entries
being used to treat correctly the edge-region close to $z=1$.
Similarly the evaluation of an operator such as $P^2$ requires
$\order{pn^2}$ operations. This therefore remains much more manageable
than the $\order{n^3}$ factor that is relevant with a non-uniform
grid, while preserving formal $p^\mathrm{th}$ order accuracy at all
values of $x$. We of course maintain the property that the evaluation
of $P \otimes q$ requires $\order{n^2/2}$ operations

We point out that the approach used here differs significantly in
philosophy from that in say \cite{Ratcliffe} in that the evolution in
$Q^2$ is performed with a Runge Kutta algorithm, rather than by a
formal analytic solution to the evolution equations expressed in terms
of a power series of $P$. This gives considerable simplicity because,
for example, the inclusion of NNLL splitting functions and 3-loop
running for $\as$ requires no extra analytical calculations.

\subsection{Testing the evolution: comparisons to tabulated global-fit
 PDF sets}
\begin{figure}[p]
  \begin{center}
    \epsfig{file=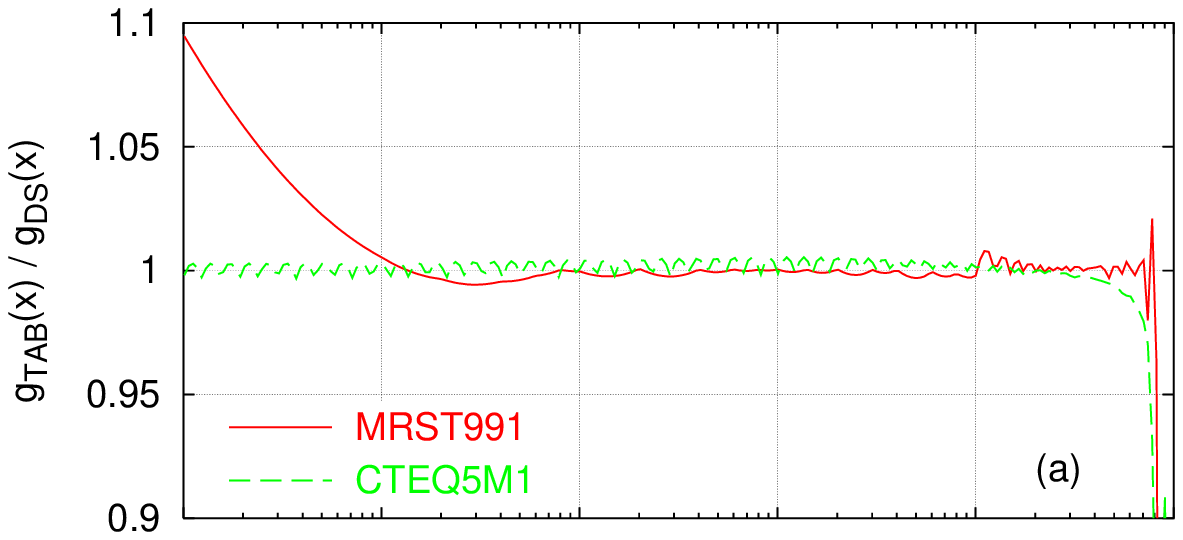,width=0.85\textwidth}\vspace{0.02\textwidth}\\
    \epsfig{file=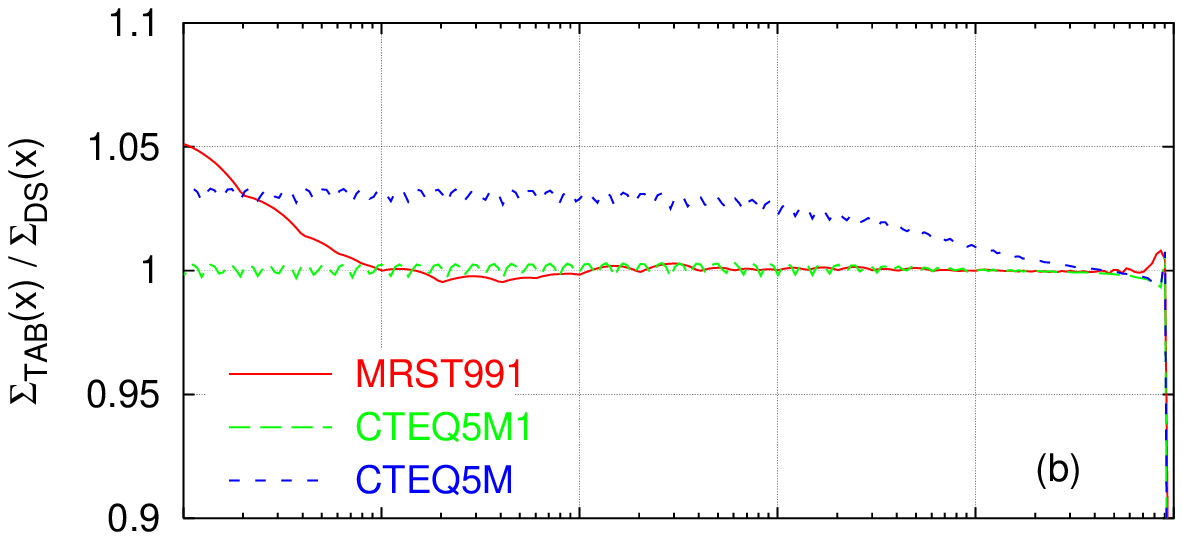,width=0.85\textwidth}\vspace{0.02\textwidth}\\
    \epsfig{file=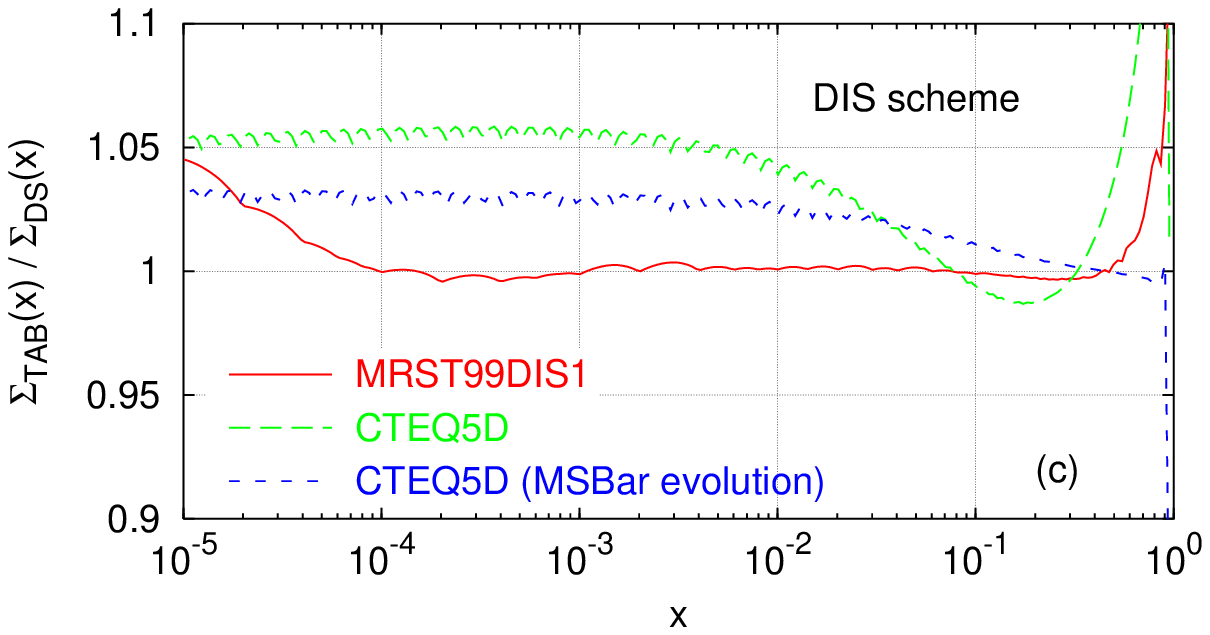,width=0.85\textwidth}\\
    \caption{Comparisons between evolution encoded in two widely used
      tabulated global-fit PDF sets  (TAB) \cite{MRST,CTEQ} and our evolution
      (DS). In all cases we 
      start with the global-fit (tabulated, interpolated)
      distribution at $10$~GeV, evolve up to $179$~GeV and then
      compare to the tabulated distribution at the higher scale; (a)
      and (b) are in the $\MSbar$ scheme, while $(c)$ is for the DIS
      scheme.}
    \label{fig:PDFcomparisons}
  \end{center}
\end{figure}

As a test of our evolution code we decided to compare to the latest
available tabulated evolution of the MRST \cite{MRST}, CTEQ
\cite{CTEQ} and GRV \cite{GRV} groups. It was our expectation that
following the work of the 1995/96 HERA workshop (see for example
\cite{Blumlein:1996rp}), the latest available tabulated distributions
would embody reasonably accurate evolution. 

The comparisons indicate
that our code is functioning properly since we generally agree with
the evolution of at
least one of the tabulated global-fit PDF sets. However in certain
situations (probably of limited practical phenomenological relevance)
we do see significant disagreement with one of MRST or CTEQ and in
what follows we give a summary of our findings. 

To carry out the comparison we adopted the following procedure. We
started with the MRST or CTEQ tabulated (interpolated in $x$)
distribution at $Q=10$~GeV and evolved it up to $179$~GeV with our
evolution routines.  We then took the ratio of the tabulated
distribution at $179$~GeV and our evolved distribution. This is
plotted for MRST and CTEQ in figure~\ref{fig:PDFcomparisons}. We have
chosen an upper limit of $179$~GeV rather than a `round number'
because it is a grid point of the MRST distributions: the
interpolation provided with the MRST99 PDF sets is linear in $Q^2$ (as
mentioned earlier) and
a choice of $Q$ in between grid points introduces errors of the order
of a couple of percent.

Figure~\ref{fig:PDFcomparisons}a shows the results for the gluon
distribution --- there is good agreement with CTEQ and MRST over
most of the range in $x$.  However at small $x \lesssim 10^{-4}$ we
find significant disagreement with MRST at a level of up to $10\%$,
whereas the agreement with CTEQ remains perfect. At large $x \gtrsim
0.5$ there is slight disagreement with the CTEQ evolution, while
agreement with MRST remains good up to $x \gtrsim 0.8$. It should be
kept in mind that in this region the gluon distribution is in any case
very small, and poorly constrained experimentally.

We note that the small irregularities of the curves arise because
slight non-smoothness of the starting distributions (a consequence of
the interpolation in $x$) gets amplified after convolutions with the
plus-distributions of the splitting functions.

Figure~\ref{fig:PDFcomparisons}b shows the analogous results for the
quark singlet distribution,
\begin{equation}
\Sigma(x) \;= \sum_{j=u,d,s,\ldots} \left( q_j(x)+\qbar_j(x)\right).
\end{equation}
The problem in the MRST distributions at small $x$ are present here
too, though they are reduced by a factor of $2$. At large $x$ the
agreement is uniformly good. We also show a comparison with the
CTEQ5M evolution which uses some numerical approximations in the NLO
evolution (see the note in the archive v3 of \cite{CTEQ}), which is
useful for reference below.

We have not explicitly shown comparisons with the GRV98 $\MSbar$
global-fit evolution \cite{GRV} --- there we find only small differences,
generally less than $1\%$, which can be ascribed to a different
convention for the treatment of higher-order terms in the evolution.
Detailed comparisons with Vogt's code, using the same convention for
the higher-order terms, gives systematic agreement to a relative
accuracy of better than $10^{-4}$ for $x < 0.8$ \cite{SalamVogt}.

We also wished to test our DIS scheme evolution,
figure~\ref{fig:PDFcomparisons}c. The DIS factorisation scheme is
defined as the scheme in which all higher-order corrections to the
$F_2$ coefficient functions are zero. Again we start with tabulated
global-fit
distributions (now in the DIS scheme) at $10$~GeV, evolve them up to
$179$~GeV 
(with DIS scheme splitting functions) and compare with the tabulated
distributions at the higher scale. Ignoring the small-$x$ problem,
with MRST there is generally good agreement.  There is a moderate
discrepancy at large $x$, however we believe that this is because of
the convention adopted by MRST to define the DIS scheme --- all their
fitting and evolution is done in the $\MSBar$ scheme, and then at each
$Q$ value they calculate the DIS-scheme distribution from the $\MSBar$
distributions. This differs from straightforward DIS scheme evolution
by NNL terms, which are largest at large $x$ because of a $\ln
(1-x)/(1-x)_+$ enhancement.

With CTEQ, the agreement is quite poor and we identify two reasons for
this. Firstly there is no updated version of the DIS scheme
distributions (\ie no equivalent to CTEQ5M1), so the numerical
inaccuracies present in CTEQ5M (see figure~\ref{fig:PDFcomparisons}b)
are present in CTEQ5D as well. Secondly it seems that the NLO
splitting functions used for the evolution are in the $\MSBar$ scheme:
if we use $\MSBar$ splitting functions for the evolution then the
resulting disagreement is identical to that due to numerical
approximations that is to be seen in CTEQ5M.

We note that the DIS scheme was not examined in the context of the
1995/96 HERA workshop studies \cite{Blumlein:1996rp} and it would
perhaps be worth performing a similar `standardisation exercise' also
in the DIS scheme.


\end{document}